\documentclass[sigconf]{acmart}
\usepackage{subfigure}
\usepackage{makecell}
\usepackage{multirow}
\AtBeginDocument{%
  \providecommand\BibTeX{{%
    \normalfont B\kern-0.5em{\scshape i\kern-0.25em b}\kern-0.8em\TeX}}}

\copyrightyear{2025} 
\acmYear{2025} 
\setcopyright{acmlicensed}\acmConference[WSDM '25]{Proceedings of the
Eighteenth ACM International Conference on Web Search and Data
Mining}{March 10--14, 2025}{Hannover, Germany}
\acmBooktitle{Proceedings of the Eighteenth ACM International Conference on
Web Search and Data Mining (WSDM '25), March 10--14, 2025, Hannover,
Germany}
\acmDOI{10.1145/3701551.3703528}
\acmISBN{979-8-4007-1329-3/25/03}

\begin{document}

\title{An Aspect Performance-aware Hypergraph Neural Network for Review-based Recommendation}

\author{Junrui Liu}
\email{liujunrui@emails.bjut.edu.cn}
\orcid{0000-0003-1632-1354}
\affiliation{%
  \institution{College of Computer Science, Beijing University of Technology}
  \city{Beijing}
  \country{China}}

\author{Tong Li}
\authornote{Corresponding author.}
\email{litong@bjut.edu.cn}
\orcid{0000-0002-8881-0037}
\affiliation{%
  \institution{College of Computer Science, Beijing University of Technology}
  \city{Beijing}
  \country{China}}

\author{Di Wu}
\email{wudi7432@gmail.com}
\orcid{0000-0003-2092-8650}
\affiliation{%
  \institution{Beijing Police College}
  \city{Beijing}
  \country{China}}

\author{Zifang Tang}
\email{zifangtang@emails.bjut.edu.cn}
\orcid{0009-0005-3485-069X}
\affiliation{%
  \institution{College of Computer Science, Beijing University of Technology}
  \city{Beijing}
  \country{China}}

\author{Yuan Fang}
\email{yfang@smu.edu.sg}
\orcid{0000-0002-4265-5289}
\affiliation{%
  \institution{School of Computing and Information Systems, Singapore Management University}
  \country{Singapore}}

\author{Zhen Yang}
\email{yangzhen@bjut.edu.cn}
\orcid{0000-0002-6058-0217}
\affiliation{%
  \institution{College of Computer Science, Beijing University of Technology}
  \city{Beijing}
  \country{China}}

\renewcommand{\shortauthors}{Junrui Liu et al.}

\begin{abstract}
Online reviews allow consumers to provide detailed feedback on various aspects of items.
Existing methods utilize these aspects to model users' fine-grained preferences for specific item features through graph neural networks.
We argue that the performance of items on different aspects is important for making precise recommendations, which has not been taken into account by existing approaches, due to lack of data.
In this paper, we propose an aspect performance-aware hypergraph neural network (APH) for the review-based recommendation, which learns the performance of items from the conflicting sentiment polarity of user reviews.
%
Specifically, APH comprehensively models the relationships among users, items, aspects, and sentiment polarity by systematically constructing an aspect hypergraph based on user reviews.
%
In addition, APH aggregates aspects representing users and items by employing an aspect performance-aware hypergraph aggregation method.
It aggregates the sentiment polarities from multiple users by jointly considering user preferences and the semantics of their sentiments, determining the weights of sentiment polarities to infer the performance of items on various aspects.
Such performances are then used as weights to aggregate neighboring aspects.
Experiments on six real-world datasets demonstrate that APH improves MSE, Precision@5, and Recall@5 by an average of $2.30\%$, $4.89\%$, and $1.60\%$ over the best baseline.
The source code and data are available at \url{https://github.com/dianziliu/APH}.
\end{abstract}

\ccsdesc[500]{Information systems~Recommender systems}

\keywords{User review, Aspect, Hypergraph, Sentiment polarity aggregation}


\maketitle

\section{Introduction}

Recommender systems have been extensively integrated into web services to enhance user experiences.
Users are encouraged to share their feelings through ratings and reviews.
Reviews contain users' opinions or sentiments about special aspects, where an aspect is a word or phrase describing a property of items and explicitly describing the characteristics of the items the user cares about~\cite{DBLP:conf/acl/HeLND17, DBLP:conf/emnlp/LiCKHCW21}.
For example, in the sentence that \textit{``Amazing sound and quality, all in one headset'',} \textit{``sound''} and \textit{``quality''} are two aspects.
As a type of important user-generated content, reviews can help recommender systems understand user preferences and item features~\cite{wang2011collaborative,daml,DBLP:conf/wsdm/ChengG022,DSRLN}.

Existing methods detect aspects in user reviews and leverage them to model users' fine-grained preferences to specific item features by graph neural networks.
For example, 
the APRE model~\cite{DBLP:conf/emnlp/LiCKHCW21} identifies the importance of aspects by considering the similarity between the aspects and their content features in reviews.
MA-GNNs~\cite{DBLP:journals/nn/ZhangXLWDLC23} constructs multiple aspect-aware user-item graphs and utilizes a routing-based fusion mechanism to allocate weights to different aspects.
RGNN~\cite{liu2021learning} regards aspects and sentiments as nodes and builds a subgraph for each user and item.
It employs a type-aware graph attention mechanism that aggregates the context information from neighboring nodes to learn the node embeddings.
A personalized graph pooling operator is proposed to learn the semantic representation for each user/item from the graph.

\begin{figure*}[htb]
    \centering
    \includegraphics[width=\linewidth]{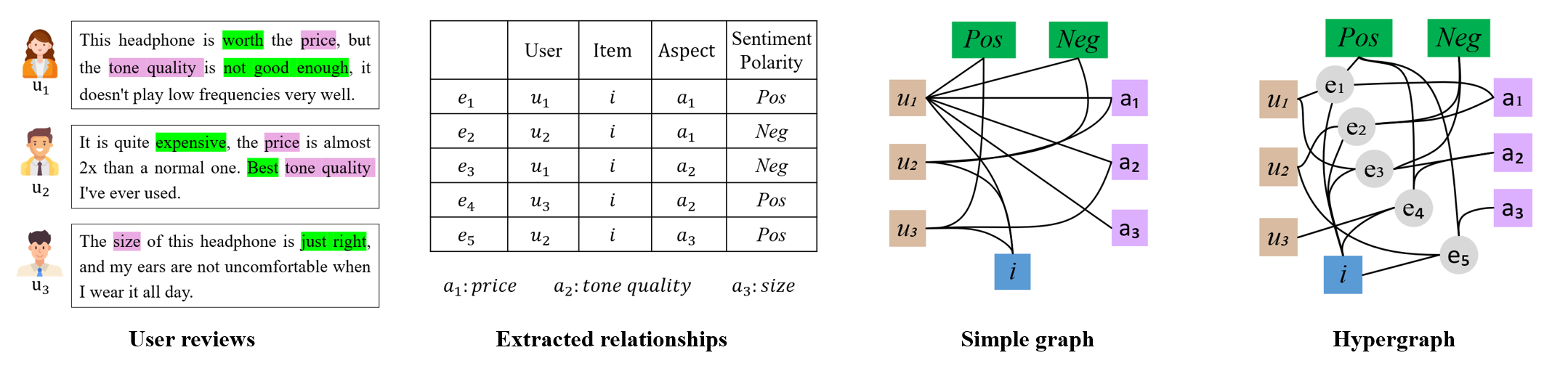}
    \caption{Hypergraph vs. simple graph. 
    There are three reviews written by three users for one headphone.
    Based on these reviews, we extract five relationships that record the users’ sentiment polarity towards various aspects of the item.
    Formally, an vertex set is $\mathcal{V}=\{u_1, u_2, u_3, i, a_1, a_2, a_3, Pos, Neg\}$ and a relationship set is $\mathcal{E}=\{e_1, e_2, e_3, e_4, e_5\}$. 
    In the simple graph, two vertices are joined together by an edge if they commonly exist in any relationship.
    This graph cannot tell us much information, like whether a user has a positive sentiment for what. 
    In the hypergraph, each hyperedge $e_n$ connects four vertices, and can completely illustrate one extracted relationship.
}
    \label{fig:hyper_vs_sim}
\end{figure*}

It is noteworthy that when users select items, they tend to prioritize item performances on various aspects. 
However, not all the performances can be directly obtained. 
This results in existing methods only considering user preferences in aspects reflected in the reviews when calculating the importance of aspects, without considering the actual performance of items in those aspects, leading to suboptimal results. 
We plan to extract and learn the performance of items on different aspects from user reviews. 
The reviews encompass users' opinions on a specific aspect of an item, and sentiment polarities in these opinions partially reflect the item's performance in that aspect. 
Nevertheless, there are conflicts in the sentiment polarities expressed by different users, which makes it challenging to accurately extract the item's performance in a given aspect from the reviews.

In reality, the conflicting sentiment polarities arise from the users' preferences in both aspects and sentiments. 
Therefore, to identify the true relationship between items and aspects, it is essential to consider user preferences when aggregating sentiment polarities.
For example, users with light tastes think a dessert has just the right amount of sweetness, while for a dessert lover, it's too light. 
Combining the feedback from multiple users and their preferences, we can assume that this dessert is on the low side of sweetness.

In this paper, we propose an aspect performance-aware hypergraph neural network (APH) for the review-based recommendation, which learns the performance of items from the conflicting sentiment polarity of user reviews.
Specifically, we first extract aspects and sentiment polarities from reviews to systematically construct an aspect-based hypergraph.
Since the representation power of simple graphs is limited and cannot express the relationships between users, items, aspects, and sentiment polarities, we choose hypergraphs with stronger representation ability to model the relationships~\cite{DBLP:journals/pami/GaoZLZDZ22}.
Figure~\ref{fig:hyper_vs_sim} compares the representation abilities of these two types of graphs.
Then, we design an aspect performance-aware hypergraph aggregation method that aggregates conflicting sentiment polarities to learn the performances of items on different aspects and treats them as weights in aggregating aspect nodes to represent the item.
The sentiment polarity contained in the reviews is somewhat subjective due to user preferences, leading to conflicting sentiment polarities between different users.
To accurately learn the performance of an item on an aspect, we aggregate multiple hyperedges related to the item and the aspect, and then assign weights to each aspect based on the performance using the attention mechanism.
%
Furthermore, users may decide whether to buy an item because of its extreme performance on an aspect or the item's overall performance.
Thus, to model the role of aspects in a user-item interaction, an aspect fusion layer aggregates the first-order aspect nodes of users and the items, respectively, to obtain the final vector representations used for prediction.
Finally, a factorization machine (FM) is used to predict.

The main contributions of this paper are as follows:
\begin{itemize}

    \item
    We propose an aspect performance-aware hypergraph neural network (APH) for the review-based recommendation, which both considers user preference for aspects and the performances of items in those aspects.

    \item
     We propose an aspect performance-aware hypergraph aggregation method that learns the performances of items on different aspects from conflicting emotional polarities.

    \item
    Experiments on six real-world datasets demonstrate that APH improves MSE, Precision@5, and Recall@5 by an average of $2.30\%$, $4.89\%$, and $1.60\%$ over the best baseline.
    
\end{itemize}


\section{Related Work}
\label{sec:related_word}
In this section, we review rating prediction tasks in recommender systems and review-based recommendation methods.
At the same time, we also discuss aspect extraction methods and graph neural networks, which are related to our methods.

\subsection{Rating Prediction}

Rating prediction is one of the most critical tasks for recommender systems and is widely known to researchers by Netflix Prize competition~\cite{DBLP:journals/computer/KorenBV09}.
Slight performance enhancements of predictions could significantly improve the recommendations~\cite{DBLP:conf/kdd/Koren08,DBLP:journals/air/ChambuaN21}.
Matrix factorization (MF) is a successful and recognized latent factor model that attracts much attention.
The variations of MF includes PMF~\cite{DBLP:conf/nips/SalakhutdinovM07}, SVD~\cite{DBLP:journals/computer/KorenBV09}, FM~\cite{DBLP:conf/icdm/Rendle10}, etc.
With the great success of deep learning in many fields, researchers tend to apply deep learning techniques to enhance the performance of rating prediction tasks, such as NCF~\cite{he2017neural}.

By the restrictions of rating sparsity, researchers employ additional information to enhance the prediction, such as user reviews~\cite{wang2011collaborative, liu2021learning}.
User reviews have a strong intrinsic correlation with user interests as they express their views on items through words. 
Analyzing the information in user reviews can provide insight into their logic and perception.
Numerous studies have attempted to enhance the prediction performance of the rating prediction approaches by employing additional information from review texts~\cite{wang2011collaborative, liu2021learning}. 

\subsection{Review-based Recommendation}

Many review-based models are proposed to improve the performance of rating prediction.
These methods can be divided into two categories.
The first category focuses on modeling the latent features in reviews.
The second category models the interactions between users and items with review-based representation.

In this first category, some methods use natural language process (NLP) techniques to extract high-level features~\cite{wang2011collaborative,cdl}.
ConvMF~\cite{convmf} and DeepCoNN~\cite{deepconn} use CNN to extract local semantics.
CARL~\cite{carl}, DAML~\cite{daml} employs the local and mutual attention of CNN to learn features from reviews, and then integrates them with the latent factor model for rating prediction.
TAERT~\cite{DBLP:journals/isci/GuoWYHCW21} uses three attention networks to model different features, i.e., word contribution, review usefulness, and latent factors.
RGNN~\cite{liu2021learning} builds a review graph for each user where nodes are words and edges are word orders.
Besides, some methods focus on fine-grained features, including explicit and implicit aspects~\cite{DBLP:journals/tois/GuanCHZZPC19,DBLP:conf/emnlp/LiCKHCW21}. 
ANR~\cite{DBLP:conf/cikm/ChinZJC18} learns aspect-based representations for the user and item by an attention-based module. 
Moreover, the co-attention mechanism is applied to the user and item importance at the aspect level. 
CAPR~\cite{DBLP:conf/sigir/LiQPQDW19} and ARPM~\cite{lai2021rating} perform aspect and sentiment analysis on textual reviews and then establish users' and items' preference feature vectors.
APRE~\cite{DBLP:conf/emnlp/LiCKHCW21} uses dependency parsing to extract explicit aspects and CNN to model user preferences based on them.
In some literature, implicit aspects-based methods are regarded as high-level features-based methods~\cite{DBLP:journals/tois/GuanCHZZPC19}.
MRCP~\cite{DBLP:journals/tois/LiuWPWWJ21} extracts word-level, review-level, and aspect-level features to represent users and items via a three-tier attention network.
SENGR~\cite{DBLP:journals/isci/ShiWGHCZH22} is a sentiment-enhanced neural graph method that incorporates the information derived from textual reviews and bipartite graphs. 


%

The second category models the interactions between users and items with review-based representation. 
D-attn~\cite{dattn} uses dual local and global attention to model word-level and review-level features.
As global attention is applied to both the user side and the item side, it learns the interaction features between the two sides.
Then the resultant factors are used for rating prediction, similar to matrix factorization. 
NARRE~\cite{narre} filters useless reviews by using the vector representing each user and item as a part of attention scores. 
HTI~\cite{wen2020hierarchical} captures interactions based on reviews by mutually propagating textual features.
Further, rather than representing users and items with static latent features, HTI dynamically identifies informative textual features at both word and review levels for each specific user-item pair. 
NRCA~\cite{DBLP:conf/sigir/Liu0XPJ20} points out two main paradigms of reviews, i.e., the document level and the review level.
It uses a cross-attention mechanism to aggregate the informative words and reviews and represent users.
DSRLN~\cite{DSRLN} extracts static and dynamic user interests by stacking attention layers that deal with sequence features and attention encoding layers that deal with of user-item interaction.
Similar to Transformers~\cite{DBLP:conf/icdm/KangM18}, DSRLN adds temporal dependencies on sequence features.



\subsection{Aspect Extraction and Sentiment Analysis}

The Aspect extraction and sentiment Analysis task aims to extract aspect term, opinion term, and their associated sentiment.
Existing methods are divided into two categories, supervised methods~\cite{DBLP:conf/acl/XuCB20,DBLP:conf/acl/ChenZFLW22}, and unsupervised methods~\cite{DBLP:conf/aaai/LiuLZKG16,DBLP:journals/ipm/DragoniFR19}.
Manually annotating data for training, which requires the hard labor of experts, is only feasible on small datasets in particular domains such as \textit{Laptop} and \textit{Restaurant}, which leads to supervised methods unsuited to our situation.
Thus, we mainly focus on unsupervised methods.
Hu and Liu~\cite{DBLP:conf/kdd/HuL04} extracted the nouns/noun phrases from sentences, and such nouns/noun phrases were labeled as aspects.
Once all aspects were selected, the nearest adjectives were extracted as potential opinion words.
Some methods tend to identify the dependency of each word and design some rules to extract aspects~\cite{DBLP:conf/aaai/LiuLZKG16, 8464692, DBLP:journals/ipm/DragoniFR19}.
Considering the impact of extracting aspect sentiment pairs on recommendation performance, we have chosen an unsupervised approach for extraction.

\subsection{Graph Neural Networks in Recommendation}


Graph neural networks (GNNs) extend deep learning techniques to process the graph data, and they are widely used in various fields~\cite{DBLP:journals/tnn/WuPCLZY21}.
Here we primarily focus on discussing GNN techniques used in reviews-based recommendation methods.

There are two main paradigms. 
One paradigm is the document-level that introduces features of reviews into a user-item graph.
RGCL~\cite{DBLP:conf/sigir/ShuaiZWSHWL22} constructs a review-aware user-item graph, where each edge feature is composed of both the user-item rating and the corresponding review semantic features. 
The feature-enhanced edges can help learn each neighbor node weight attentively for user and item representation learning.
The other paradigm is word-level, i.e., regard words in reviews as graph nodes and then aggregates them to represent users/items.
These methods mainly use graph attention networks to represent nodes by aggregating their neighbor nodes\cite{DBLP:journals/corr/BrunaZSL13,DBLP:journals/corr/abs-1710-10903}.
The attention mechanism determines the weights of neighbor nodes.
Li et al.~\cite{DBLP:conf/emnlp/LiCKHCW21} identify the importance of aspects by attention mechanism that regards the content features of aspects in reviews as the query to calculate the weights.
The DualGCN model~\cite{DBLP:journals/kbs/ShiWHZCZH22} regards aspects and sentiments as nodes in aspect graphs and adopts sum pooling to represent users and items.
RGNN~\cite{liu2021learning} builds a review graph for each user where nodes are words, and edges are word orders.
It uses a type-aware method, which regards the combination of the type of edges and neighbor nodes as keys in the attention mechanism, to aggregate the information of neighboring words effectively.
The MA-GNNs model~\cite{DBLP:journals/nn/ZhangXLWDLC23} predefines four aspects and constructs multiple aspect-aware user-item graphs, regarding the aspect-based sentiment as the edge.
It utilizes a routing-based fusion mechanism to allocate weights to different aspects, realizing the dynamic fusion of aspect preferences.

\section{Preliminaries}

The hypergraph is a generalization of the graph~\cite{DBLP:journals/pami/GaoZLZDZ22}.
Different from the graph, an edge in the hypergraph, called hyperedge, is a subset of all vertices in the hypergraph. 
A hypergraph is defined as $\mathcal{G}=(\mathcal{V},\mathcal{E}, \phi)$, which includes vertex set $\mathcal{V}$, a hyperedge set $\mathcal{E}$, and a node type mapping function $\phi: \mathcal{V} \rightarrow \mathcal{T}$. 
Here $\mathcal{T}$ denotes the sets of predefined node types.
A hypergraph $\mathcal{G}$ can be described by an $|\mathcal{V}| \cdot |\mathcal{E}|$ incidence matrix $\textbf{H}$, whose entries are defined as $\textbf{H}(v,e)=\left\{  \begin{aligned}
        \begin{array}{ll}
            1 & if~~v \in e \\
            0 & otherwise 
        \end{array}
\end{aligned} \right.$.
    

\section{Method}
\begin{figure*}[htb]
    \centering
    \includegraphics[width=\textwidth]{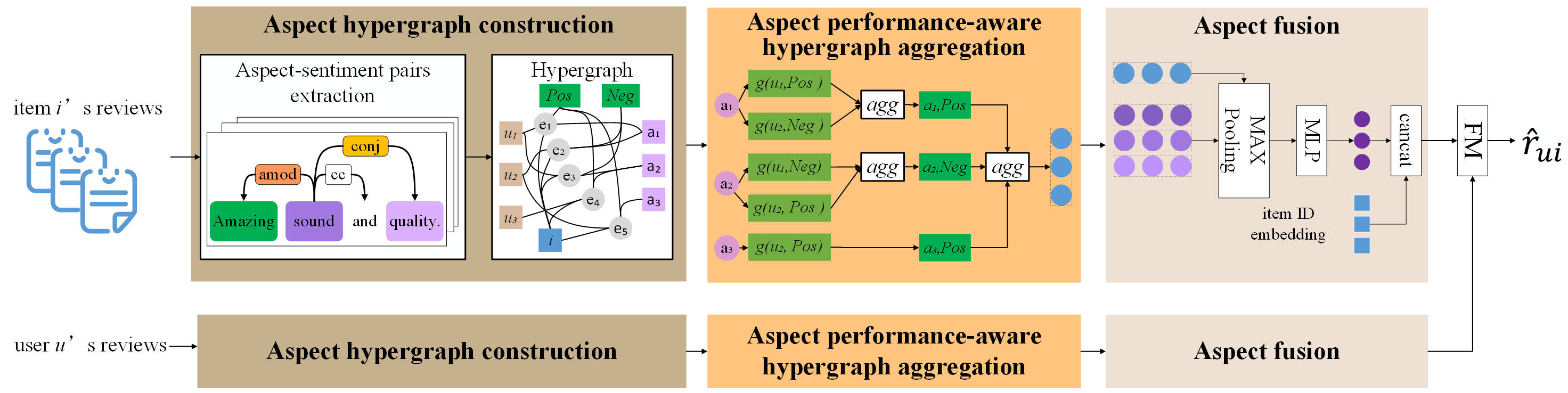}
    \caption{Framework of APH. It first extracts aspects and user sentiments from reviews to construct a hypergraph. Then, to learn the true relationship between an item and an aspect from conflicting user sentiments, APH considers user preferences to identify the weight of their sentiments. 
    Likewise, we use a similar way to calculate the aspect-based user representations.
    Finally, APH fuses items, neighbor aspect nodes, and their ID embeddings to make predictions.}
    \label{fig:overview}
    \vspace{-0.3cm}
\end{figure*}

To model user preferences for aspects and the performance of items in these aspects, APH extracts aspect-sentiment pairs from reviews.
%
The overview of APH is shown in Figure~\ref{fig:overview}.
Specifically, APH has four steps for recommendation predictions: aspect hypergraph construction, aspect performance-aware hypergraph aggregation, aspect fusion, and prediction.
The aspect hypergraph construction extracts aspect-sentiment pairs from reviews and then constructs the graph.
Considering the user preference for aspects and the performance of items in those aspects, an aspect performance-aware hypergraph aggregation layer is designed to learn the aspect weights based on user sentiments.
To learn more about the role of aspects in user-item interactions, we design the aspect fusion layer that aggregates neighboring aspects of users/items to represent users and items.
The prediction layer fuses users and items to make predictions.

\subsection{Aspect Hypergraph Construction}
\label{sec:Aspect_graph_constractor}


We aim to extract the explicit aspect, a word or phrase that describes the property of items~\cite{DBLP:conf/acl/HeLND17, DBLP:conf/emnlp/LiCKHCW21}.
Through explicit aspects, recommender systems learn the fine-grained preferences that describe the special properties of items that users are interested in.
Most existing supervised methods are trained based on \textit{Laptop} and \textit{Restaurant} datasets, which cannot fully meet our scenarios.
Thus, we utilize a rule-based unsupervised method to extract aspects and sentiments from reviews~\cite{DBLP:conf/emnlp/LiCKHCW21}. 







The rule-based unsupervised method considers three dependency relations\footnote{We use spaCy(\url{https://spacy.io}) to extract the syntactic relations between the words.} sequentially used to extract candidate aspect-sentiment pairs, including \textit{amod}, \textit{nsubj+acomp}, and \textit{dobj}.
Table~\ref{tab:rule_based_aspect} summarizes how to extract aspects and sentiments based on dependency relations.
On the one hand, some nouns directly describe the properties of items.
They are considered potential aspects and are modified by adjectives with two types of dependency relations, \textit{amod} and \textit{nsubj+acomp}. 
The pairs of nouns and the modifying adjectives compose the AS-pair candidates. 
For instance, in the sentence \textit{``Amazing sound and quality, all in one headset''}, the nouns \textit{``sound''} and \textit{``quality''} are two aspects of an item, and the user thinks the item in these aspects are amazing, which is a positive sentiment.
On the other hand, the predicate and the object in a sentence describe the function of items.
Thus, the combinations of predicates and objects are considered potential aspects by \textit{dobj} dependency relation.
In a combination, the predicate is regarded as the sentiment as it is usually with users' emotions and the whole combination is regarded as the aspect.
For example, in the sentence \textit{``If you're recording vocals this will eliminate the pops'',} the verb \textit{``eliminate''} and noun \textit{``pops''} construct a \textit{dobj} relation.
From this sentence, we know that the mentioned device is used to filter pops, and the word \textit{``eliminate''} includes the user's emotions.
Finally, to simplify the complexity of modeling, we use a sentiment analysis tool to judge the sentiment polarity of sentiment words~\cite{DBLP:journals/ipm/DragoniFR19}. 
Three kinds of positive emotions, neutral emotions, and negative emotions were extracted from sentiment words~\footnote{We use the Opinion Lexicon, which is available at \url{https://www.cs.uic.edu/~lzhang3/programs/OpinionLexicon.html}.}.
For more details please refer to the Appendix.


\begin{table}[]
    \centering
    \caption{Extracting aspects based on dependency relations}
    \vspace{-0.3cm}
    \begin{tabular}{cccc}
    \hline
        No. & Dependency relations & Aspect & Sentiment \\
        \hline
        1 &   Adj. (x) $\leftarrow$ amod $-$ Noun (y)  & y &x\\
        2 &  Noun (x) $\leftarrow$nsubj$-$ Linking Verb (y) $-$  & \multirow{2}{*}{x} & \multirow{2}{*}{z}\\
        &     \multicolumn{1}{r}{$-$ acomp $\rightarrow$ Adj. (z) }    &  & \\
        3&  Verb (x)  $-$ dobj $\rightarrow$ Noun (y) & (x,y)  & x \\
        \hline
    \end{tabular}
    \vspace{-0.5cm}
    \label{tab:rule_based_aspect}
\end{table}

By the aspect-sentiment extraction method, we extract aspect-sentiment pairs for each review.
After adding the context of users and items, we obtain quadruples.
The form of a quadruple is $(u,i,a,s)$, where $u$ is a user, $i$ is an item, $a$ is an aspect, $s$ is the sentiment polarity.
Since hypergraphs have stronger expressive power than simple graphs, we construct hypergraphs based on quadruple~\cite{DBLP:conf/nips/ZhouHS06}.
Figure~\ref{fig:hyper_vs_sim} compares the expressive power of hypergraphs and simple graphs.
In our graph, the set of predefined node types, $\mathcal{T}$,  contains four types, i.e., user $\mathcal{U}$, item $\mathcal{I}$, aspect $\mathcal{A}$, and sentiment polarity $\mathcal{S}$.
Each hyperedge contains a user, an item, an aspect, and a sentiment polarity that is the same as the quadruple.

\subsection{Aspect Performance-aware Hypergraph Aggregation\label{sec:saa}}


Formally, $\mathcal{A}_i$ is the aspect set associated with item $i$.
Existing methods~\cite{DBLP:conf/emnlp/LiCKHCW21,DBLP:journals/kbs/ShiWHZCZH22,DBLP:journals/nn/ZhangXLWDLC23} aggregate aspects to represent items, is as follows:
\begin{equation}
\begin{aligned}
    \textbf{x}_i&=f(\mathcal{A}_i)\\
    &=\sum_{a \in \mathcal{A}_i} w(\textbf{x}_a) \cdot \textbf{x}_a,
\end{aligned}
\label{eq:f1}
\end{equation}
where $w(x)$ is a weight function and is usually a softmax function, $\textbf{x}_a$ is the embedding vector of the aspect $a$.
However, Equation~(\ref{eq:f1}) overlooks the user's demands for aspects and the performance of items in these aspects.
It should be re-write as 
\begin{equation}
\begin{aligned}
    \textbf{x}_i&=f(\mathcal{A}_i)\\
    &=\sum_{a \in \mathcal{A}_i} w(\textbf{x}_a,p_{i}(a)) \cdot \textbf{x}_a,
\end{aligned}
\label{eq:f2}
\end{equation}
where $p_{i}(a)$ is a performance metric function to denote the performance of item $i$ on the aspect $a$.

Learning the performance metric function $q$ is challenging in practical situations.
We lack data about the performance of items on various aspects but have a lot of conflicting user sentiments about them.
Users' sentiment not only indicates that they care about certain aspects but also reflects whether the performance of the items aligns with their needs in this aspect.
However, sentiment polarities expressed by different users are conflicting.
The subjective feelings encompassed in user sentiments may not accurately describe the relationships between different aspects of items.
Therefore, the method should consider user preferences in the aggregation process to identify the true relationships between items and aspects.
$\mathcal{U}_i$ is the user set that rated the item $i$, and $\mathcal{S}_i$ is the sequence list of users' sentiments on aspects.
Our aggregation representation of an item is:
\begin{equation}
\begin{aligned}
    \textbf{x}_i&=f(\mathcal{A}_i,q_{i}(\mathcal{S}_i,\mathcal{U}_i))\\
    &=\sum_{a \in \mathcal{A}_i }{
    \sum_{u,s \in \mathcal{U}_i,\mathcal{S}_i} {w(\textbf{x}_a,q_{i}(\textbf{x}_u,\textbf{x}_s)) \cdot \textbf{x}_a}},
\end{aligned}
\label{eq:f3}
\end{equation}
where the function $q$ is used to acquire the aspect performance metric function from user preferences and sentiments.
It aggregates the sentiment polarities of multiple users by jointly considering their preferences and the semantic meaning, thereby determining the weights of the sentiment polarities.

We extend the aforementioned approach to the hypergraph we have constructed.
$\mathcal{E}_i$ is the set of hyperedges that connected with an item $i$, and $\mathcal{E}_{i,a} $ it the subset with an aspect $a$.
For each hyperedge $e \in \mathcal{E}_{i,a}$, its weight is:
\begin{equation}
\begin{aligned}
    w(e)&=w(u,i,a,s) \\
    &=\frac{
    exp[\pi(\textbf{x}_i,q_{i}(\textbf{x}_u,\textbf{x}_s),\textbf{x}_a)]
    }{
    \sum_{e^{\prime} \in \mathcal{E}_{i}} {exp[\pi(\textbf{x}_i,q_{i}(\textbf{x}_{u^{\prime}},\textbf{x}_{s^{\prime}}),\textbf{x}_a)]}
    },
\end{aligned}
\label{eq:weight_edge}
\end{equation}
where $\textbf{x}_u,~\textbf{x}_i,~\textbf{x}_a,~\textbf{x}_s \in \mathbb{R}^{1 \times d_1}$ are the input embeddings of nodes.
Note that the calculation range of the softmax function is $\mathcal{E}_i$ instead of $\mathcal{E}_{i, a}$.
The former can obtain the weight of each edge according to the performance difference of different aspects, while the latter makes the weight of each aspect equal to 1 after aggregating the edges related to the aspect.

Considering the nonlinear relationship between sentiment features and user preferences, we use MLP as the implementation of the function $q_i$.
The result is denoted as $\textbf{x}_q \in \mathbb{R}^{1 \times d_1}$ and is following:
\begin{equation}
\textbf{x}_q=q_i(\textbf{x}_u,\textbf{x}_s)=MLP(\textbf{x}_u,\textbf{x}_s),
\end{equation}
$\pi(\textbf{x}_i,\textbf{x}_q,\textbf{x}_a)$ is implemented by the following relational attention mechanism:
\begin{equation}
    \pi(\textbf{x}_i,\textbf{x}_q,\textbf{x}_a)=LeakyRelu[(\textbf{x}_i \textbf{W}_1) (\textbf{x}_q \textbf{W}_2+\textbf{x}_a \textbf{W}_3)]
    \label{eq:f_pi}
\end{equation}
$\textbf{W}_1, \textbf{W}_2, \textbf{W}_3 \in \mathbb{R}^{d_1 \times d_2}$ are three different weight matrices used for the linear transformations of the node edge embedding, and $d_2$ denotes the dimension of the hidden space of the graph representation learning layers.
Finally, we aggregate all aspects in $\mathcal{E}_{i}$ to represent the item $i$, as following
\begin{equation}
    \hat{\textbf{x}}_{i}=\sum_{\mathcal{E}_{i,a} \in \mathcal{E}_{i}} \sum_{e \in \mathcal{E}_{i,a} }w(e) \textbf{x}_a\textbf{W}_4,
\label{eq:xlh}
\end{equation}
where $\textbf{W}_4 \in  \mathbb{R}^{d_1 \times d_2}$ denotes the transform matrix.
Likewise, we use a similar way to calculate the aggregation representation $\hat{\textbf{x}}_{u}$ of a user $u$.

\subsection{Aspect Fusion}

Users may decide whether to buy an item because of its extreme performance in an aspect or the item's overall performance.
To learn more about the role of aspects in user-item interactions, we have designed the aspect fusion layer that aggregates neighboring aspects of users/items to represent users and items.
For an item $i$, their aspect neighbors $\mathcal{A}_i$'s feature matrix is denoted as $\textbf{X}_{i} \in  \mathbb{R}^{|\mathcal{A}_{i}| \times d_2}$.
The rows in ${\textbf{X}}_i$ is ranked under Equation~(\ref{eq:f_pi}).
We use the max-pooling and MLP to generate a representation as follows,
\begin{equation}
\hat{\textbf{g}}_i=max^{d_2}_{t=1} {\textbf{X}}_i(:,t), 
\end{equation}
\begin{equation}
    \textbf{g}_i=ReLU(\hat{\textbf{g}}_i \textbf{W}_6+b_6),
\end{equation}
where $t$ is a hyperparameter that determines the number of aspects that are aggregated, $\textbf{W}_6 \in \mathcal{R}^{d_2 \times d_2 }$ and $b_6 \in \mathcal{R}^{1 \times d_2}$ the weight matrix and bias vector. 
Then, we concatenate the non-linear transform of the item embedding $\textbf{x}_i \in \mathcal{R}^{1 \times d_1}$ after a MLP layer and the representation of sentiment-aware aggregation to generate the aspect-aware representation of the item $i$ $\textbf{q}_i$  as follows,
\begin{equation}
    \textbf{m}_i=ReLU(\hat{\textbf{x}}_{i}~\textbf{W}_7 +b_7),
\end{equation}
\begin{equation}
\textbf{y}_i=\textbf{m}_i \oplus \textbf{g}_i,
\end{equation} 
where $\oplus$ is the concatenating operation, $\textbf{W}_7 \in \mathcal{R}^{d_1 \times d_2}$ and
$b_7 \in \mathcal{R}^{1 \times d_2}$ are the weight matrix and bias vector. 
Similarly, we can get the semantic representation of a user $u$ as $\textbf{y}_u$.

\subsection{Prediction}

An FM layer is used to predict final scores~\cite{DBLP:conf/icdm/Rendle10}.
It considers the higher-order interactions between the user and item fine-grained features. 
Specifically, we first concatenate the user and item representations as $\textbf{z} = \textbf{y}_u \oplus \textbf{y}_i$, and the prediction $r_{ui}$ is defined as follows
\begin{equation}
\hat{r}_{ui}=b_0+b_u+b_i+\textbf{z}~\textbf{w}^T+\sum^{d^{\prime}}_{i=1}\sum^{d^{\prime}}_{j=i+1}<\textbf{v}_i,\textbf{v}_j>\textbf{z}_i~\textbf{z}_j,
\end{equation}
where $b_0$, $b_u$, and $b_i$ are the global bias, user bias, and item bias, respectively. 
$\textbf{w} \in \mathcal{R}^{1 \times d^{\prime}}$ is the coefficient vector, and $d^{\prime} = 4 \times d_2$. $\textbf{v}_i,\textbf{v}_j \in \mathcal{R}^{1\times d^{\prime}}$ are the latent factors associated with
$i$-th and $j$-th dimension of $\textbf{z}$. 
$z_i$ is the value of the $i$-th dimension of $\textbf{z}$.
The model parameters $\Phi$ of APH can be learned by solving the following optimization problem,
\begin{equation}
\mathcal{L}= \min  \frac{1}{|D|}\sum_{u,i \in D} {(r_{ui}-\hat{r}_{ui})}^2+ \lambda {||\Phi ||}^2_F,
\end{equation}
where $\lambda$ is the regularization parameter, and $D$ denotes the set of user-item pairs used to update the model parameters.

\subsection{Time complexity}

APH mainly has three parts, hypergraph aggregation (HA), aspect fusion (AF), and FM for prediction.
For each prediction, the time complexity of HA is $\mathcal{O}_{HA}(|\mathcal{E}_{i}|d^3)$, where $|\mathcal{E}_{i}|$ is the number of hyperedges item $i$ connected,
that of AF is $\mathcal{O}_{AF}(d^2)$, and that of FM is  $\mathcal{O}_{AF}(d^2)$.
Summary, the time complexity of APH is $\mathcal{O}(|\mathcal{E}_{i}|d^3)$ for each user-item pair.

\section{Experiment}
In this section, we do a series of experiments to identify the performance of our method.
We describe our experimental setup and show comparison results with different baselines.
We further use an ablation study to identify the effect of each part in APH.
Finally, we analyze the extracted aspects.
For more experiments, such as hyperparameter analysis, please refer to the Appendix.

\subsection{Dataset}
The experiments are performed on the Amazon review and Yelp datasets, which have been widely used for recommendation research~\cite{daml,liu2021learning}. 
For the Amazon review dataset, we choose the following 5-core review subsets for evaluation: Musical Instruments, Office Products, Toys and Games, Video Games, and Beauty (respectively denoted by Music, Office, Toys, Games, and Beauty). 
For the Yelp dataset, we only keep the users and items that have at least 10 reviews for experiments. 
Table~\ref{tab:dataset} summarizes the details of these experimental datasets.
We discuss the extract aspects in section~\ref{sec:ana_asepct}.

\begin{table}[]
    \centering
    \caption{The statistics of the experimental datasets.}
    \vspace{-0.3cm}
    \begin{tabular}{ccccc}
    \hline
         Dataset & \#Users &\#Items &\#Ratings/Reviews & \#Density \\
         \hline
         Music  &1,429  &900     &10,261         &0.80\%    \\
         Office &4,905  &2,420   &53,228         &0.45\%    \\
         Toys   &19,412 & 11,924 &167,597        &0.07\%       \\
         Games  &24,303 & 10,672 &231,577        &0.09\%       \\
         Beauty &22,363 & 12,101 &198,502        &0.07\%       \\
         Yelp   &26,084 & 65,786 &3,519,533      &0.04\%       \\
        \hline
    \end{tabular}
    \label{tab:dataset}
\end{table}

\subsection{Baselines}

We compare APH with three types of baselines.
Traditional rating-based methods include:{PMF}~\cite{DBLP:conf/nips/SalakhutdinovM07}, {SVD}++~\cite{DBLP:journals/computer/KorenBV09}.
Review-based methods include: {CDL}~\cite{cdl}, {DeepCoNN}~\cite{deepconn}, {NARRE}~\cite{narre}, {ANR}~\cite{DBLP:conf/cikm/ChinZJC18}, {CARL}~\cite{carl}, {DAML}~\cite{daml}, and {DSRLN}~\cite{DSRLN}. Aspect-based methods include: {NRCA}~\cite{DBLP:conf/sigir/Liu0XPJ20}, {MA-GNNs}~\cite{DBLP:journals/nn/ZhangXLWDLC23}, and {RGNN}~\cite{liu2021learning}.
These methods have been discussed in Section~\ref{sec:related_word}.
As MA-GNNs is trained by pairwise loss, we only compared it with NDCG.

\subsection{Setup}

For each dataset, we randomly choose $20\%$ of the user-item review pairs (denoted by $D_{test}$) for evaluating the model performance in the testing phase, and the remaining $80\%$ of the review pairs (denoted by $D_{train}$) are used in the training phase. 

In recommender systems, there are two common tasks: rating prediction and click-through rate prediction.
Thus, to evaluate the performance of our method in the rating prediction task, we apply the typically used Mean Square Error (MSE) and Normalized Discounted Cumulative Gain (NDCG), which has been widely used in previous studies~\cite{deepconn,carl,liu2021learning};
we use Precision(Pre) and Recall(Rec) for click-through rate prediction to evaluate the Top-K performance.

\subsection{Performance Comparison}


\subsubsection{Model performance on rating prediction task}
\begin{table}[]
\caption{The performances of different recommendation methods evaluated by MSE. The best results are in bold faces and the second-best results are underlined. * indicates that the standard deviation of the results of the five times is smaller than 0.001. }
\vspace{-0.3cm}
\centering
\scalebox{0.9}
{\begin{tabular}{lcccccc}
\hline
Dataset         &Music     &Office   &Toys         &Games    &Beauty 	   &Yelp       \\
\hline
PMF             &1.8783	   &0.9635	 &1.6091	   &1.5260	 &2.7077	   &1.4217    \\
SVD++           &0.7952	   &0.7213	 &0.8276	   &1.2081	 &1.2129	   &1.2973    \\
\hline
CDL             &1.2987	   &0.8763	 &1.2479	   &1.6002	 &1.7726	   &1.4042    \\
DCN        &0.7909	   &0.7315	 &0.8073	   &1.1234	 &1.2210	   &1.2719    \\
NARRE           &0.7688	   &0.7266	 &0.7912	   &1.1120	 &1.1997	   &1.2675    \\
CARL	        &0.7632	   &0.7193	 &0.8248	   &1.1308	 &1.2250	   &1.3199    \\
DAML	        &0.7401	   &0.7164	 &0.7909	   &1.1086	 &1.2175	   &1.2700    \\
NRCA	        &0.7658	   &0.7343	 &0.8100	   &1.1259	 &1.2034	   &1.2721    \\
DSRLN	        &0.7538	   &0.7131	 &0.8141	   &1.1205	 &1.1951	   &\underline{1.1655}    \\
\hline
ANR             &0.7825	   &0.7237	 &0.7974	   &1.1038	 &1.2021	   &1.2708    \\
RGNN	        &\underline{0.7319}	   &\underline{0.7125}	 &\textbf{0.7786}	   &\underline{1.0996}	 &\underline{1.1885}	   &1.2645    \\
\hline
APH	        &\textbf{0.6795*}&0\textbf{.6884*}&\underline{0.7859*}&\textbf{1.0829}	 &\textbf{1.1757*}&\textbf{1.1467*}    \\
\hline
\end{tabular}
}
\label{tab:per_comp}        
\vspace{-0.5cm}
\end{table}


%
The MSE and NDCG results of the performance comparison are shown in Table~\ref{tab:per_comp} and Table~\ref{tab:res:ndcg}, respectively.
We mark the best results in bold faces and the second-best results are underlined.
APH achieves the best results compared with other baselines in five datasets and the second-best results in the remaining one.
On average, APH improves MSE by $2.30\%$ compared to the best baseline.
Nevertheless, our approach has achieved an improvement in NDCG.
These results show that APH can effectively improve prediction performance by modeling the performance of items in aspects.
Explicit aspects describe the fine-grained preferences of users and explain the characteristics of the items that the user cares about.
APH models fine-grained preferences from explicit aspect-sentiment pairs to enhance prediction performance.
To drop out subjective feelings in user sentiments and identify the true relationships between items and aspects, APH designs an aspect performance-aware aggregation layer that separates the user preferences from the sentiment.
Thus, APH effectively improves recommendation performance.
In the Toys dataset, RGNN performs a better MSE result than APH.
We observe that the most frequent aspects in the Toy dataset reflect the characteristics of the user, followed by the characteristics of the item, which is different from other datasets. 
This situation could amplify the variance of the prediction error.
In summary, APH models user preferences for aspects and the performance of items in these aspects, enhancing recommendation performance.

\begin{table}[]
\caption{The performances of different recommendation methods evaluated by NDCG@10. The best results are in bold faces and the second-best results are underlined. 
* indicates that the standard deviation of the results of the five times is smaller than 0.001.}
\vspace{-0.3cm}
\scalebox{0.9}
{\begin{tabular}{lcccccc}
\hline
Dataset         &Music     &Office   &Toys         &Games    &Beauty 	   &Yelp       \\
\hline
DCN   & 0.977 & 0.973 & 0.975 & 0.971 & 0.966  & 0.941 \\
NARRE & 0.978 & 0.976 & 0.981 & 0.968 & 0.971  & 0.957 \\
CARL  & 0.980  & 0.978 & 0.978 & 0.969 & 0.966  & 0.943 \\
DAML  & \underline{0.982} & 0.978 & 0.979 & \textbf{0.979} & 0.967  & 0.958 \\
DSRLN & 0.781 & 0.974 & 0.977 & \textbf{0.979} & 0.967  & 0.948 \\
\hline
MA-GNNs & 0.979 & 0.973 & 0.975 & 0.966 & 0.965  & 0.933 \\
RGNN  & \underline{0.982} & \underline{0.983} & \underline{0.982} & 0.976 & \underline{0.973}  & \underline{0.963} \\
\hline
APH   & \textbf{0.988*} & \textbf{0.986*} & \textbf{0.983*} & \underline{0.977*} & \textbf{0.974*}  & \textbf{0.965*} \\
\hline
\end{tabular}
}
\label{tab:res:ndcg}
\vspace{-0.3cm}
\end{table}

\subsubsection{Model performance on click-through rate prediction}

\begin{table*}[]
\centering
\caption{The performances of different recommendation methods evaluated by P@5 and R@5. The best results are in bold faces and the second-best results are underlined.
* and $\ddag$ ~indicate that the Standard Deviation of the results of the five times is smaller than 0.001 and 0.002, respectively.}
\vspace{-0.2cm}
\begin{tabular}{lllllllllllll}
\hline
& \multicolumn{2}{c}{Music}                          & \multicolumn{2}{c}{Office}                          & \multicolumn{2}{c}{Toys}                          & \multicolumn{2}{c}{Games}                          & \multicolumn{2}{c}{Beauty}                      & \multicolumn{2}{c}{Yelp}                       \\
      & \multicolumn{1}{c}{Pre@5} & \multicolumn{1}{c}{Rec@5} & \multicolumn{1}{c}{Pre@5} & \multicolumn{1}{c}{Rec@5} & \multicolumn{1}{c}{Pre@5} & \multicolumn{1}{c}{Rec@5} & \multicolumn{1}{c}{Pre@5} & \multicolumn{1}{c}{Rec@5} & \multicolumn{1}{c}{Pre@5}     & \multicolumn{1}{c}{Rec@5} & \multicolumn{1}{c}{Pre@5}    & \multicolumn{1}{c}{Rec@5} \\
\hline
DCN   & 0.2327    & 0.6818    & 0.2555    & 0.5953    & 0.2408    & 0.6228    & 0.2561    & 0.6355    &    0.2876           & 0.7024          &  0.3238            &  0.5985         \\
NARRE & 0.2502    & 0.6603    & 0.3265    & 0.7361    & 0.0105    & 0.0341    & 0.2053    & 0.4984    & 0.1545        & 0.4094    & 0.3976       & 0.5907    \\
DAML  & 0.2515    & 0.7019    & 0.3158    & 0.6796    & 0.2517    & 0.6638    & 0.2598    & 0.6622    & 0.2227        & 0.5911    &    0.3861          &    0.6138       \\
RGNN  & 0.2690    & 0.7453    & 0.3229    & 0.6967    & 0.2874    & 0.7599    & \underline{0.2809}    & \underline{0.7164}    & {0.2985}       & 0.7387    &  0.3824            &0.6592           \\
DSRLN & \underline{0.2721}    & \underline{0.7518}    & \underline{0.3386}    & \underline{0.7386}    & \underline{0.2873}    & \underline{0.7503}    & {0.2673}    & {0.7131}    &    \underline{0.3044}           &    \underline{0.7642}       &     \underline{0.4278}           &  \textbf{0.7248}         \\
APH   & \textbf{0.2730*}    & \textbf{0.7566$^{\ddag}$}    & \textbf{0.3461*}    & \textbf{0.7433$^{\ddag}$}    & \textbf{0.2985 *}   & \textbf{0.7614*}    & \textbf{0.3263*}    & \textbf{0.7890*}    & \textbf{0.3158*}        & \textbf{0.7753*}    &\textbf{0.4407*}             & \underline{0.6996*}  \\

\hline
\end{tabular}
\label{tab:Pre_Rec_res}
\vspace{-0.2cm}
\end{table*}

For CTR prediction, we use cross-entropy loss to train all models and add a sigmoid layer as the activation function.
The ratio of negative sampling is $4$, i.e., we sample $4$ negative items from unobserved items for each positive item.
Other settings are the same as those for rating prediction.
The numerical results on all the benchmark datasets are displayed in Table~\ref{tab:Pre_Rec_res}.
APH achieves the best results compared with other baselines in six datasets.
Compared to the base baseline, APH achieves an average improvement $4.89\%$ on Pre@5 and $1.60\%$ on Rec@5.
For the CTR task, APH can more effectively distinguish the difference between positive and negative items than baseline, by learning the item's performance in certain aspects.
The design of APH helps recommender systems recommend more accurately.

\subsection{Ablation Study}


\begin{table}[]
\caption{MSE results of ablation study.}
\vspace{-0.3cm}
\scalebox{0.9}{
\begin{tabular}{ccccccc}
\hline
Dataset  & Music  & Office  & Toys    & Games     & Beauty  & Yelp       \\
\hline
APH(MAX) & 0.7006    & 0.6951    & 0.7972    & 1.0879   & 1.1773    & 1.1913\\
APH(MEAN) & 0.6933	 &0.7010     &0.7918     &1.0799     &1.1820	&1.1755   \\
APH(-AF)  & 0.6873  &0.7068     &0.8040     &1.0958     &1.1899     &1.1869  \\
APH(-FM)  & 0.8173    & 0.7196    & 0.8228    & 1.1052 & 1.1999  & 1.1714\\
APH       & \textbf{0.6795}    & \textbf{0.6884}    & \textbf{0.7859}    & \textbf{1.0829} & \textbf{1.1757}  & \textbf{1.1467}\\
\hline
\end{tabular}
}
\label{tab:Ablation_Study}
\vspace{-0.2cm}
\end{table}

To investigate the importance of each component of APH, we consider the following variants of APH for experiments:
\begin{itemize}
    \item APH(MAX/MEAN) dropouts the aspect performance-aware aggregation layer and uses max/mean pooling instead. 

    \item APH(-AF) dropouts the aspect fusion layer.
    \item APH(-FM) uses the dot function to predict ratings.
\end{itemize}

The experiment results are shown in Table~\ref{tab:Ablation_Study}.
It determines that the aspect performance-aware aggregation layer and the aspect fusion layer positively impact performance.
User sentiments contain users' subjective feelings and do not effectively describe the relationships between different aspects of items.
Therefore, APH proposes an aspect performance-aware aggregation layer, which learns the performance of items in aspects from user sentiments.
We use the max and mean pooling layers instead of the aggregation layer, leading the model to fail to determine the performances of items on the aspects.
Despite introducing the aspect performance-aware aggregation layer effectively identifying the importance of aspects in the hypergraph, achieving satisfactory results during the prediction phase still relies on successfully matching user demands with item performance.
To learn more about the role of aspects in user-item interactions, we have designed the aspect fusion layer that aggregates neighboring aspects of users/items to represent users and items.
From the results, we can find that the aspect fusion layer successfully aggregates aspects to represent users and items and matches them for predictions.
In a word, these two layers are important for our method.


\subsection{Analysis of Extracted Aspect}
\label{sec:ana_asepct}

\begin{table}[]
    \centering
    \caption{The statistics of explicit aspects in various datasets.}
    \vspace{-0.3cm}
    \begin{tabular}{lcc}
    \hline
Dataset & \# Aspect & \# Quadruple  \\
\hline
Music   & 601       & 38,898    \\
Office  & 3,092      & 393,038   \\
Toys    & 4,809     & 776,819   \\
Games   & 11,656     & 2,439,534  \\
Beauty  & 4,868      & 866,835   \\
Yelp    & 43,904     & 20,857,681 \\ 
\hline
    \end{tabular}
    \label{tab:sta_asp}
     \vspace{-0.3cm}
\end{table}

\begin{figure}
    \centering
    \includegraphics[width=5cm]{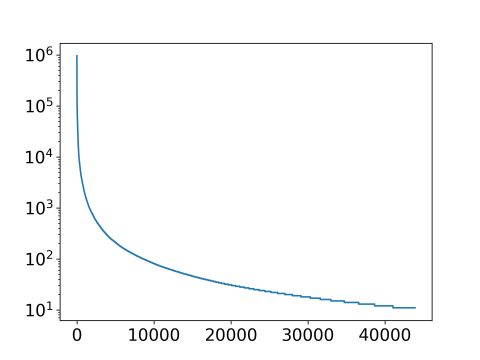}
    \caption{Aspect distribution in the Yelp dataset, which is similar to the other five datasets. }
    \label{fig:Aspect_distribution}
    \vspace{-0.5cm}
\end{figure}

Our method aggregates explicit aspects to represent users and items and fuses them to make predictions.
The aspects are extracted by an unsupervised method, discussed in section~\ref{sec:Aspect_graph_constractor}.
In this subsection, we show the statistics of explicit aspects in Table~\ref{tab:sta_asp}.
The number of aspects is smaller than that of items, and the number of quadruples is bigger than that of ratings.
These characteristics can help to reduce the size of the parameter space.
We also give the distribution of aspects in Figure~\ref{fig:Aspect_distribution}.
Most distributions are long-tail distributions.
This paper focuses on the impact of aspects, so the impact of distributions will be studied in the future.
Top-10 explicit aspects in various datasets give an overview of the quality of extracted aspects, which are shown in Table~\ref{tab:top10_asp}.
We can see that the extracted aspects include some normal terms, like ``quality'', ``color'', ``price'', and some special terms like ``amp'', ``skin'', etc.
It determines that our method can effectively extract explicit aspects from reviews.
In some situations, our method faces trouble and regards ``son'' and ``daughter'' as aspects that reduce model performance, which leads our method to play the second performance in this dataset.



\begin{table}[]
\caption{Top-10 explicit aspects in various datasets.}
\vspace{-0.3cm}
\scalebox{0.9}{
\begin{tabular}{|c|c|c|c|c|c|}
\hline
Music   & Office  & Toys     & Games   & Beauty     & Yelp       \\
\hline
quality & quality & toy      & back    & hair       & place      \\
guitar  & mark    & kid      & graphic & product    & food       \\
draw    & color   & part     & way     & scent      & service    \\
good    & printer & daughter & thing   & skin       & staff      \\
price   & product & boy      & quality & color      & restaurant \\
one     & price   & quality  & level   & have\_hair & selection  \\
wheel   & part    & back     & point   & price      & price      \\
thing   & paper   & one      & part    & have\_skin & portion    \\
base    & thing   & fit      & work    & face       & experience \\
amp     & size    & thing    & control & smell      & sauce     \\
\hline
\end{tabular}
}
\label{tab:top10_asp}
\vspace{-0.3cm}
\end{table}

\subsection{Case study}


\begin{figure}
    \centering
    \subfigure[Extracted aspect quadruples.]{\includegraphics[width=0.95\linewidth]{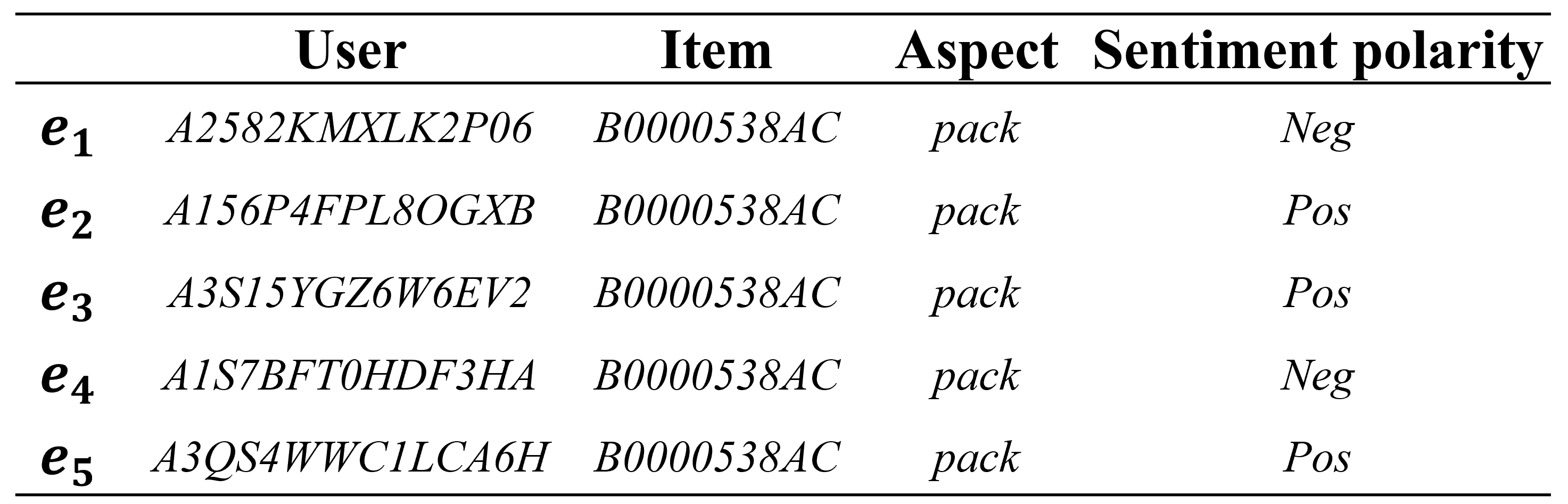}}
    \vfill
        \subfigure[The subgraph of the item.]{\includegraphics[width=0.38\linewidth]{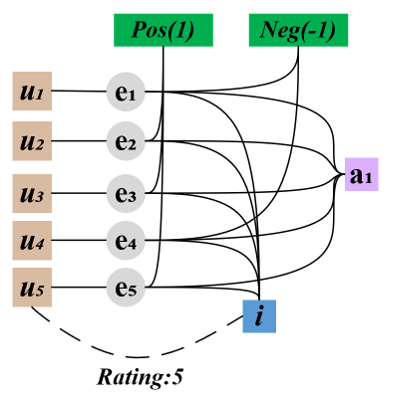}}
            \subfigure[Intermediate results of the model.]{\includegraphics[width=0.57\linewidth]{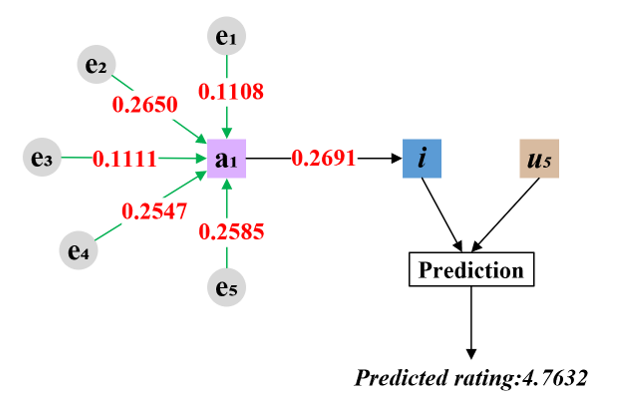}}
    \vspace{-0.3cm}
    \caption{A case study. We show the extracted aspect quadruples of an item ``$B0000538AC$'' and an aspect ``pack''; then we build the subgraph; we also show the attention scores calculated by Equation ~\eqref{eq:weight_edge}, and the final predicted rating.}
    \label{fig:case_study}
    \vspace{-0.5cm}
\end{figure}

In addition, we also conduct a case study to explore whether APH learns the performance of items on an aspect.
We randomly select an item ``$B0000538AC$'' from the Office dataset that has conflicting sentiment polarities from different users.
Figure~\ref{fig:case_study} visualizes related aspect quadruples, the subgraph of the item, the attention scores of user sentiment polarities calculated by Equation ~\ref{eq:weight_edge}, and the final predictive rating.
APH considers the item's performance of the ``pact'' aspect to be $0.2691$, higher than the mean of the sentiment polarities($Neg=-1, Pos=1$).
When the aggregation aspects represent an item, the aspect performance-aware hypergraph aggregation layer calculates the performance of the item in aspects based on the user's sentiment polarities, making the aggregation results more accurate.

\section{Conclusion}

Due to the performances of items on aspects being unavailable in datasets, existing methods only consider user preferences in aspects reflected in the reviews when aggregating aspects, and do not consider the actual performance of items in those aspects, leading to suboptimal results. 
We argue that the performances can be extracted and learned from user reviews. 
To this end, this paper proposes an aspect performance-aware hypergraph neural network for recommender systems, which considers user preference for aspects and the performance of items in those aspects when calculating their importance.
We extract aspect-sentiment pairs from reviews and then construct an aspect-based hypergraph.
Subsequently, we design a method that incorporates user preferences in aspect sentiment pairs to aid in aggregating conflicting sentiment features and learn the item's performance in each aspect.
An aspect fusion layer respectively combines aspects with users and items, modeling the role that aspects play in the interaction between users and items.
Experiments on six real-world datasets demonstrate that the predictions of APH significantly outperform baselines. 
In future work, we plan to extract aspect categories to enhance the connectivity of aspect graphs.

%
%


\section*{Acknowledgments}
This work is partially supported by the National Key R\&D Program of China (No.2022YFB3103100), 
and the Major Research Plan of the National Natural Science Foundation of China (No.92167102).

\bibliographystyle{ACM-Reference-Format}
\bibliography{main}

\newpage
\appendix

\section{The Details of Aspect Hypergraph Construction}
\label{sec:Aspect_graph_constractor_app}
Although Section 4.1 is self-explanatory, we would like to explain more about the AS-pair generation process in this section.

\begin{table*}
\caption{Using dependency to extract aspect-sentiment pairs.}\label{tab:feature_rule}
\begin{tabular}{ccc}
\hline
{No.} & \textbf{Dependency} & \textbf{Example}\\
\midrule
\makecell{\textbf{AF1}}  &  
\begin{minipage}[b]{0.5\columnwidth}
		\raisebox{-.5\height}{\includegraphics[width=\linewidth]{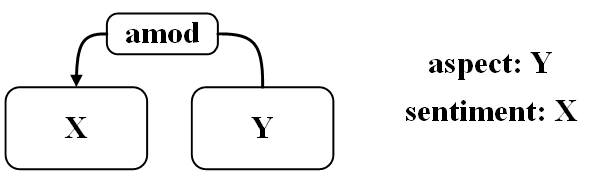}} 
	\end{minipage}  &  
   \begin{minipage}[b]{1\columnwidth}
		\raisebox{-.5\height}{\includegraphics[width=\linewidth]{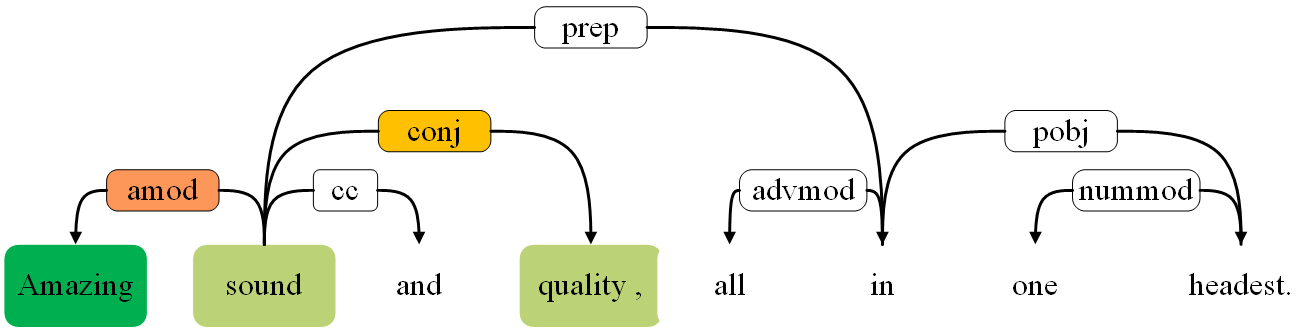}} 
	\end{minipage}       \\

\makecell{\textbf{AF2}}  &  
   \begin{minipage}[b]{0.7\columnwidth}
		\raisebox{-.5\height}{\includegraphics[width=\linewidth]{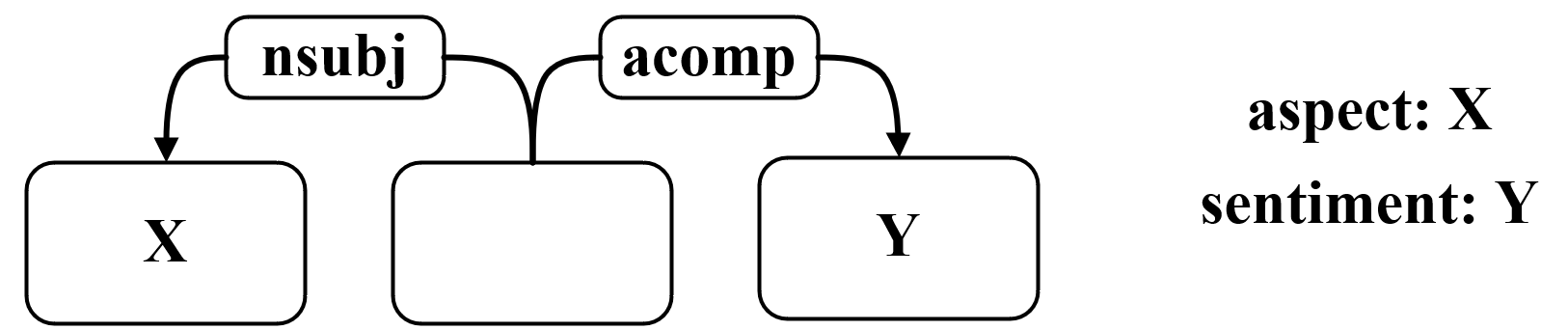}} 
	\end{minipage} 
& 
   \begin{minipage}[b]{\columnwidth}
		\raisebox{-.5\height}{\includegraphics[width=0.9\linewidth]{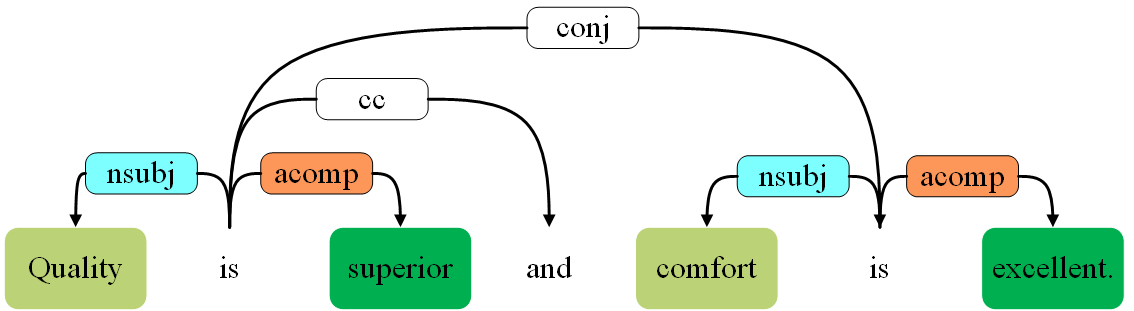}} 
	\end{minipage}       \\

\makecell{\textbf{AF3}}  & 
   \begin{minipage}[b]{0.5\columnwidth}
		\raisebox{-.5\height}{\includegraphics[width=\linewidth]{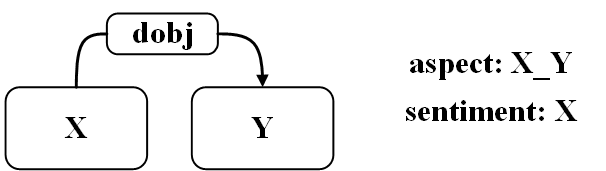}} 
	\end{minipage} & 
   \begin{minipage}[b]{\columnwidth}
		\raisebox{-.5\height}{\includegraphics[width=\linewidth]{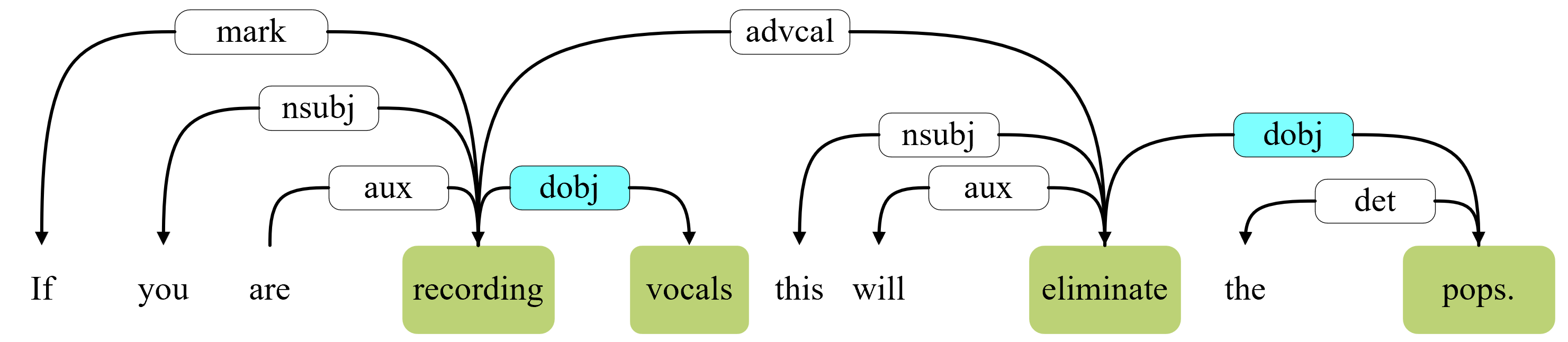}} 
	\end{minipage}       \\
 \bottomrule
\end{tabular}
\end{table*}

\subsection{Rule-based extraction}
The rule-based extraction method considers three dependency relations\footnote{We use spaCy(\url{https://spacy.io/}) to extract the syntactic relations between the words.} sequentially used to extract candidate aspect-sentiment pairs, including \textit{amod}, \textit{nsubj+acomp}, and \textit{dobj}, shown in Table~\ref{tab:feature_rule}.

On the one hand, some nouns directly describe the properties of items.
They are considered potential aspects and are modified by adjectives with two types of dependency relations, \textit{amod} and \textit{nsubj+acomp}. 
The pairs of nouns and the modifying adjectives compose the AS-pair candidates. 
For instance, in the sentence \textit{Amazing sound and quality, all in one headest}, the adjective \textit{amazing} and the nouns \textit{sound} compose a \textit{amod} relationship.
The nouns \textit{sound} and \textit{quality} are two aspects of an item, and the user thinks these aspects of the item are \textit{amazing}, which is a positive sentiment.
Thus, we extract adjectives as sentiment words and the related nouns as aspects.
Sometimes, users tend to comment an aspect with a complete structure, such as \textit{Quality is superior and comfort is excellent }, where \textit{quality} is the subject and \textit{superior} is the object.
The dependency relation \textit{nsubj+acomp} is suitable for this situation.
On the other hand, the predicate and the object in a sentence describe the function of items.
Thus, the combinations of predicates and objects are considered potential aspects by \textit{dobj} dependency relation.
In a combination, the predicate is regarded as the sentiment as it is usually with users' emotions and the whole combination is regarded as the aspect.
For example, in the sentence \textit{If you're recording vocals this will eliminate the pops,} the verb \textit{eliminate} and noun \textit{pops} construct a \textit{dobj} relation.
From this sentence, we know that the mentioned device is used to filter pops, and the word \textit{eliminate} includes the user's emotions.


\subsection{Filtering}


To improve the quality of aspects, we merge synonyms and filter out aspects that are low in frequency.
After that, we judge the polarities from sentiment words.
\begin{itemize}
    \item Synonyms merging.
A user may use two words with similar meanings for a particular method. 
Therefore, it is necessary to merge words with the same meaning.
Merging synonyms closes the relationship between users and items obtained through aspect-sentiment pairs modeling and reduces the learning complexity of subsequent models.
We collect the synonyms for each aspect and regard the most frequent synonym as the new aspect.

\item Low-frequently aspects filtering.
Filtering by setting a threshold can filter out part of the noise.
In this paper, we set the threshold $c_t=10$ to filter out the noise pairs.

\item Sentiment polarity extraction. To simplify the modeling complexity, we use a sentiment analysis tool to judge the sentiment polarity of sentiment words. 
Three kinds of positive emotions, neutral emotions, and negative emotions were extracted from sentiment words~\footnote{We use the Opinion Lexicon, which is available at \url{https://www.cs.uic.edu/~lzhang3/programs/OpinionLexicon.html}.}.

\end{itemize}


\section{Additional Experiments}
\label{sec:add_experiment}
This section exhibits additional content regarding the experiments, such as a detailed experimental setup, the instructions to reproduce the baselines and our model, supplemental experimental results, and another case study.
We hope the critical content helps readers gain deeper insight into the performance of the proposed framework.

\subsection{Baseline}

We compare APH with three types of baselines.
Traditional rating-based methods include:
\begin{itemize}
\item \textbf{PMF}~\cite{DBLP:conf/nips/SalakhutdinovM07} is the probabilistic matrix factorization model, which is a classical collaborative filtering-based rating prediction method.
\item \textbf{SVD}++~\cite{DBLP:journals/computer/KorenBV09} is a classic matrix factorization method that exploits both the user’s explicit preferences on items and the influences of the user’s historical items on the target item.
\end{itemize}
Review-based methods include:
\begin{itemize}
\item \textbf{CDL}~\cite{cdl} is a hierarchical Bayesian model that employs SDAE for learning features from the content information and collaborative filtering for modeling the rating behaviors.

\item \textbf{DeepCoNN} (DCN)~\cite{deepconn} contains two parallel networks, which focus on modeling the user behaviors and learning the item properties from the review data.
\item \textbf{NARRE}~\cite{narre} uses an attention mechanism to model the importance of reviews and a neural regression model with review-level explanations for rating prediction.

\item \textbf{CARL}~\cite{carl} is a context-aware representation learning model for rating prediction, which uses convolution operation and attention mechanism for review-based feature learning and factorization machine for modeling high-order feature interactions.

\item \textbf{DAML}~\cite{daml} employs CNN with local and mutual attention mechanisms to learn the review features and improve the interpretability of the recommendation model.

\item \textbf{NRCA}~\cite{DBLP:conf/sigir/Liu0XPJ20} uses a review encoder to learn the review representation and a user/item encoder with a personalized attention mechanism to learn user/item representations from reviews.

\item \textbf{DSRLN}~\cite{DSRLN} extracts static and dynamic user interests by stacking attention layers that deal with sequence features and attention encoding layers that deal with of user-item interaction.
\end{itemize}
Aspect-based methods include:
\begin{itemize}
\item \textbf{ANR}~\cite{DBLP:conf/cikm/ChinZJC18} is an aspect-based neural recommendation model that learns aspect-based representations for the user and item by an attention-based module. Moreover, the co-attention mechanism is applied to the user and item importance at the aspect level. 
\item \textbf{MA-GNNs}~\cite{DBLP:journals/nn/ZhangXLWDLC23} predefines four aspects and constructs multiple aspect-aware user-item graphs, regarding the aspect-based sentiment as the edge. As it is trained by pairwise loss, we only compared it with NDCG.

\item \textbf{RGNN}~\cite{liu2021learning} builds a review graph for each user where nodes are words and edges are word orders.
It uses a type-aware graph attention network to summarize graph information and a personalized graph pooling operator to capture important aspects.

\end{itemize}

\subsection{Parameter Sensitivity Study}

This subsection explores the effect of learning rate, regularization parameter, and embedding dimension.

\subsubsection{The effect of learning rate and regularization parameter}
\begin{figure}
    \centering
    \subfigure[Music]{\includegraphics[width=4cm]{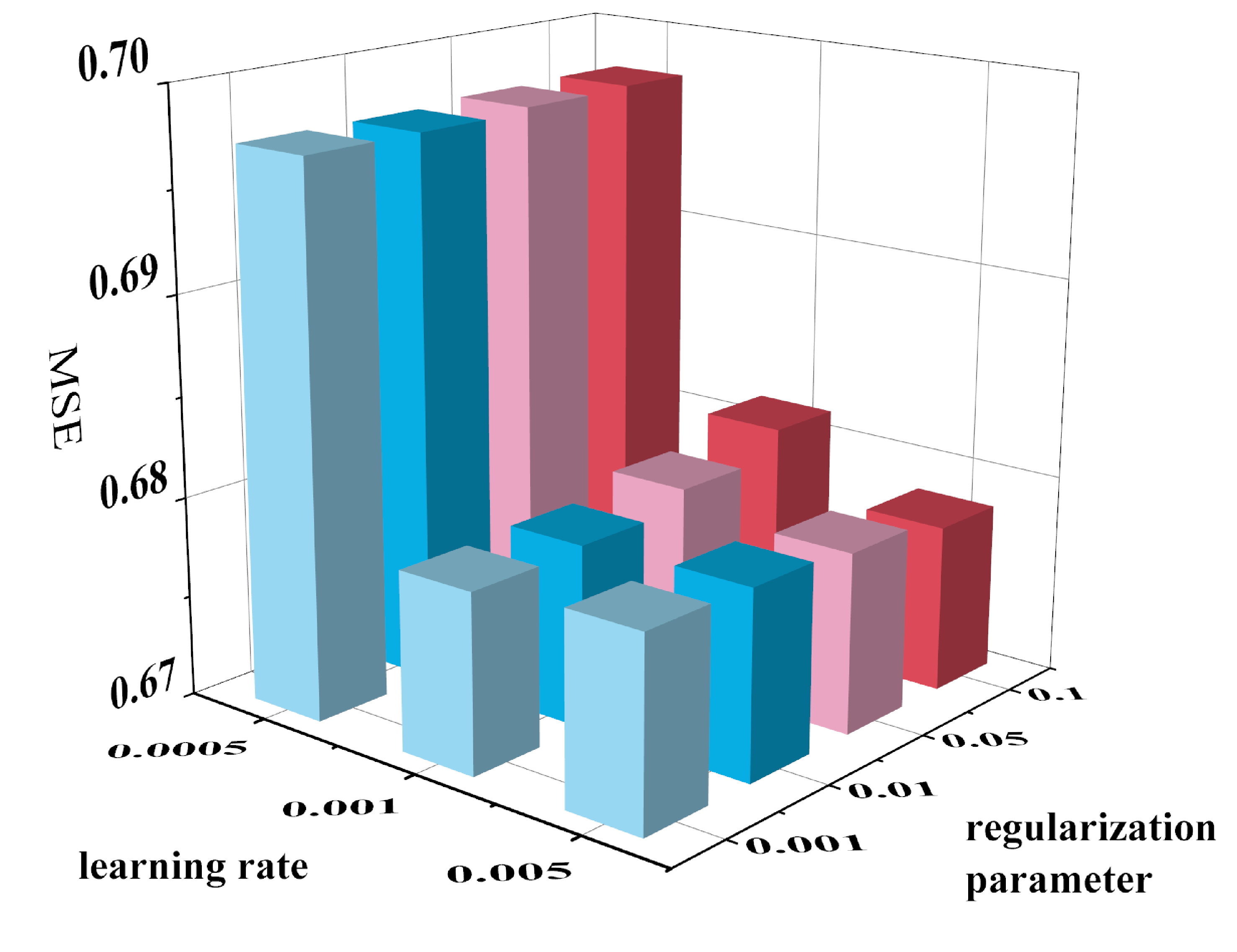}}
    \subfigure[Office]{\includegraphics[width=4cm]{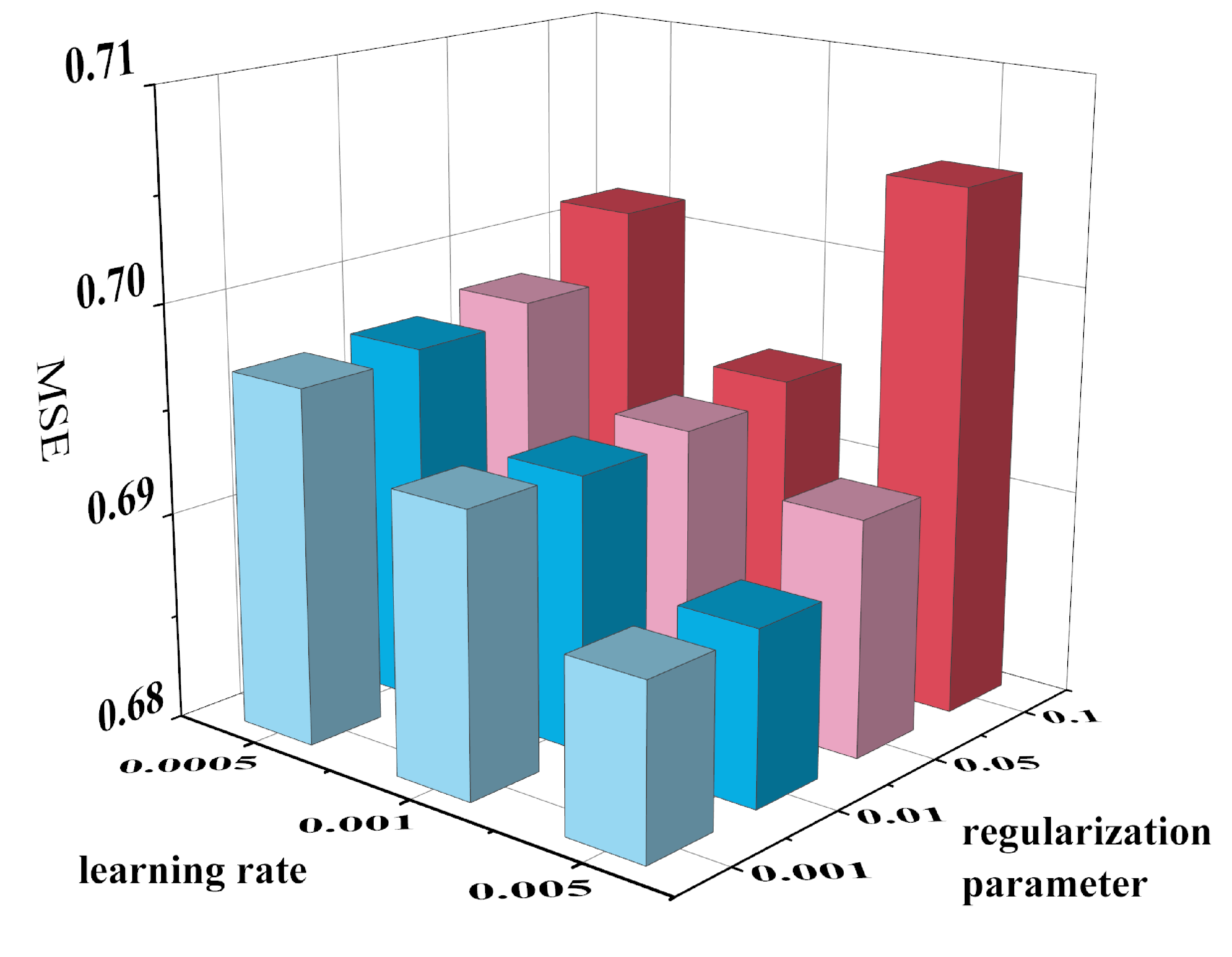}}
    \vfill
    \subfigure[Toys]{\includegraphics[width=4cm]{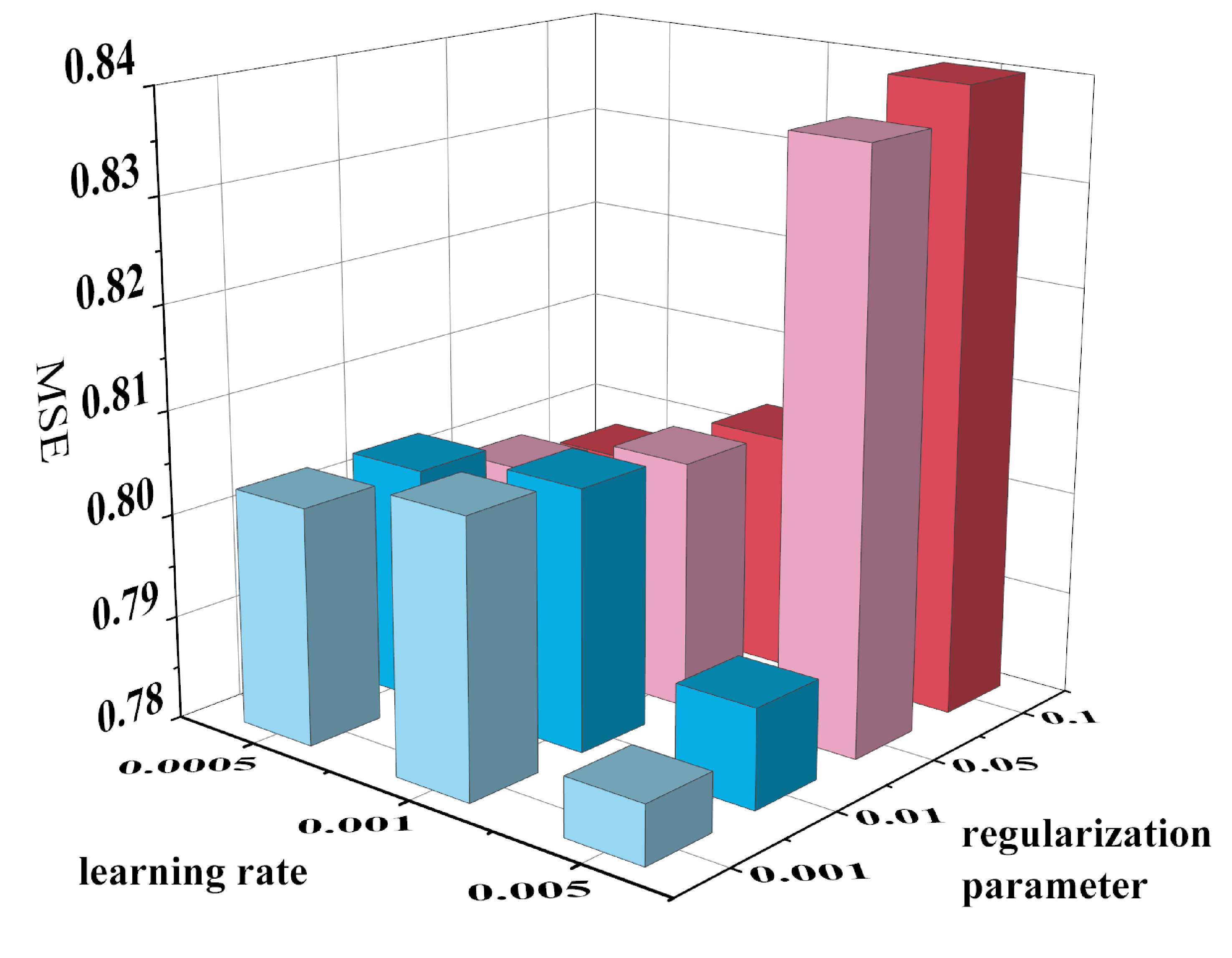}}
    \subfigure[Games]{\includegraphics[width=4cm]{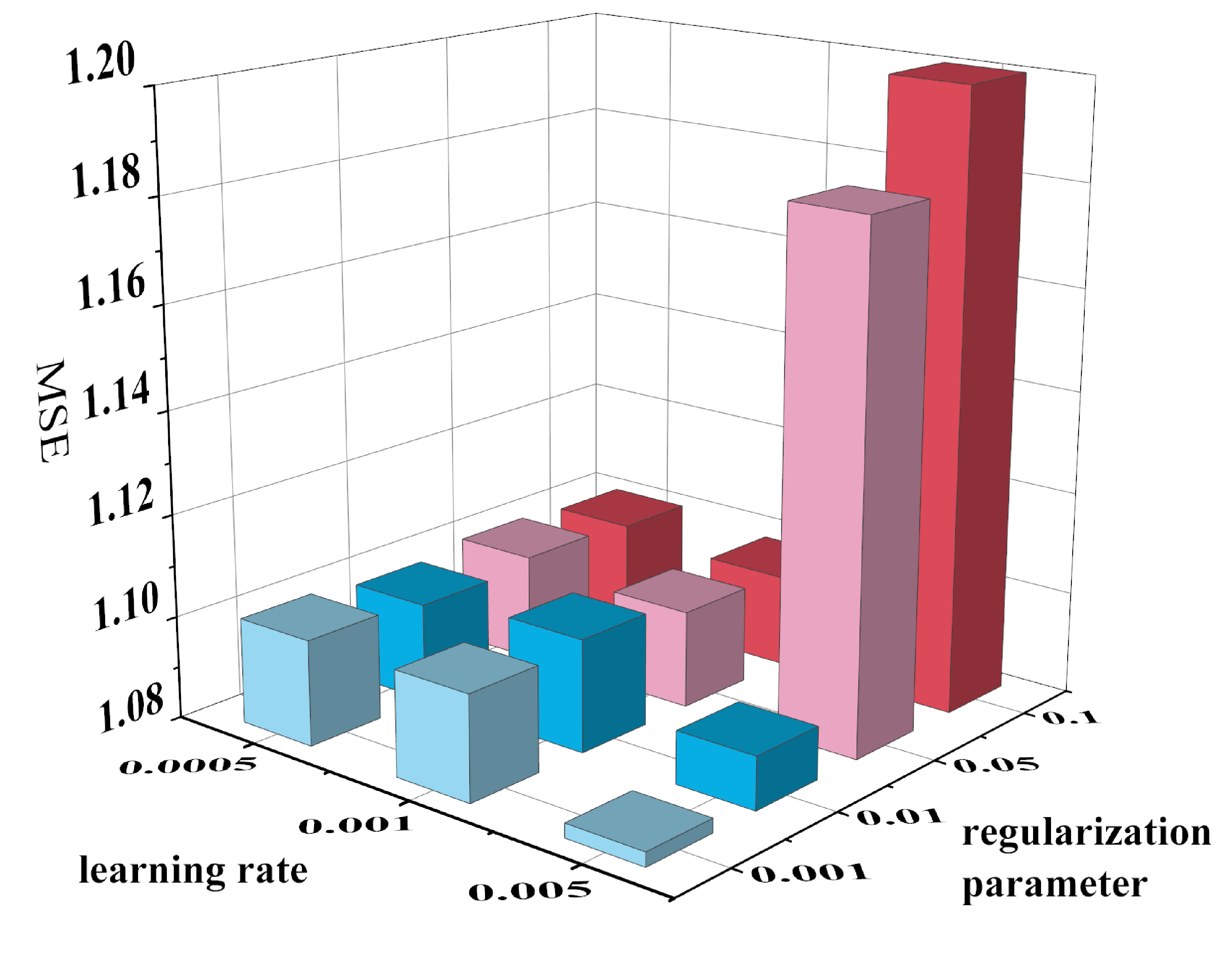}}
    \vfill
    \subfigure[Beauty]{\includegraphics[width=4cm]{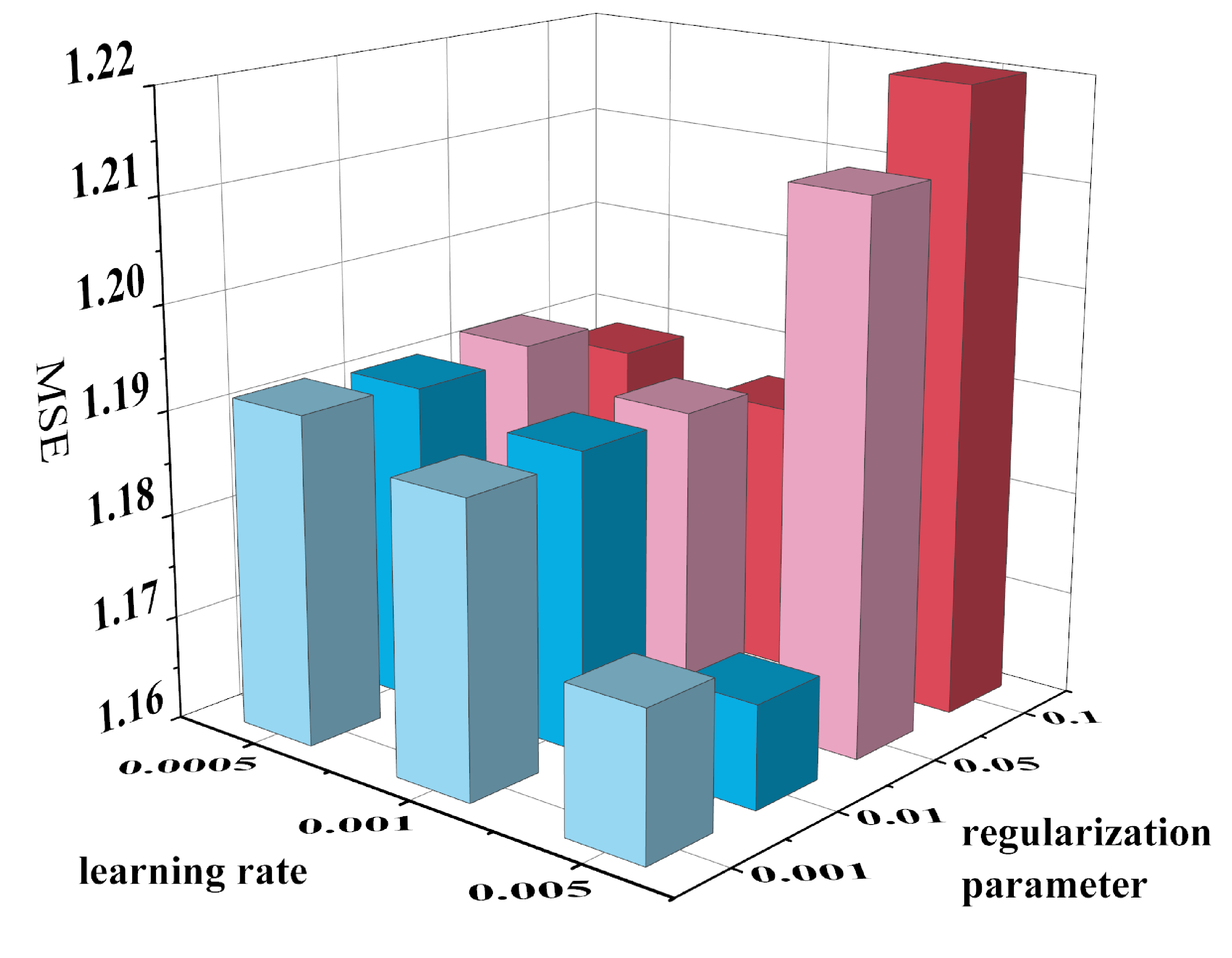}}
    \subfigure[Yelp]{\includegraphics[width=4cm]{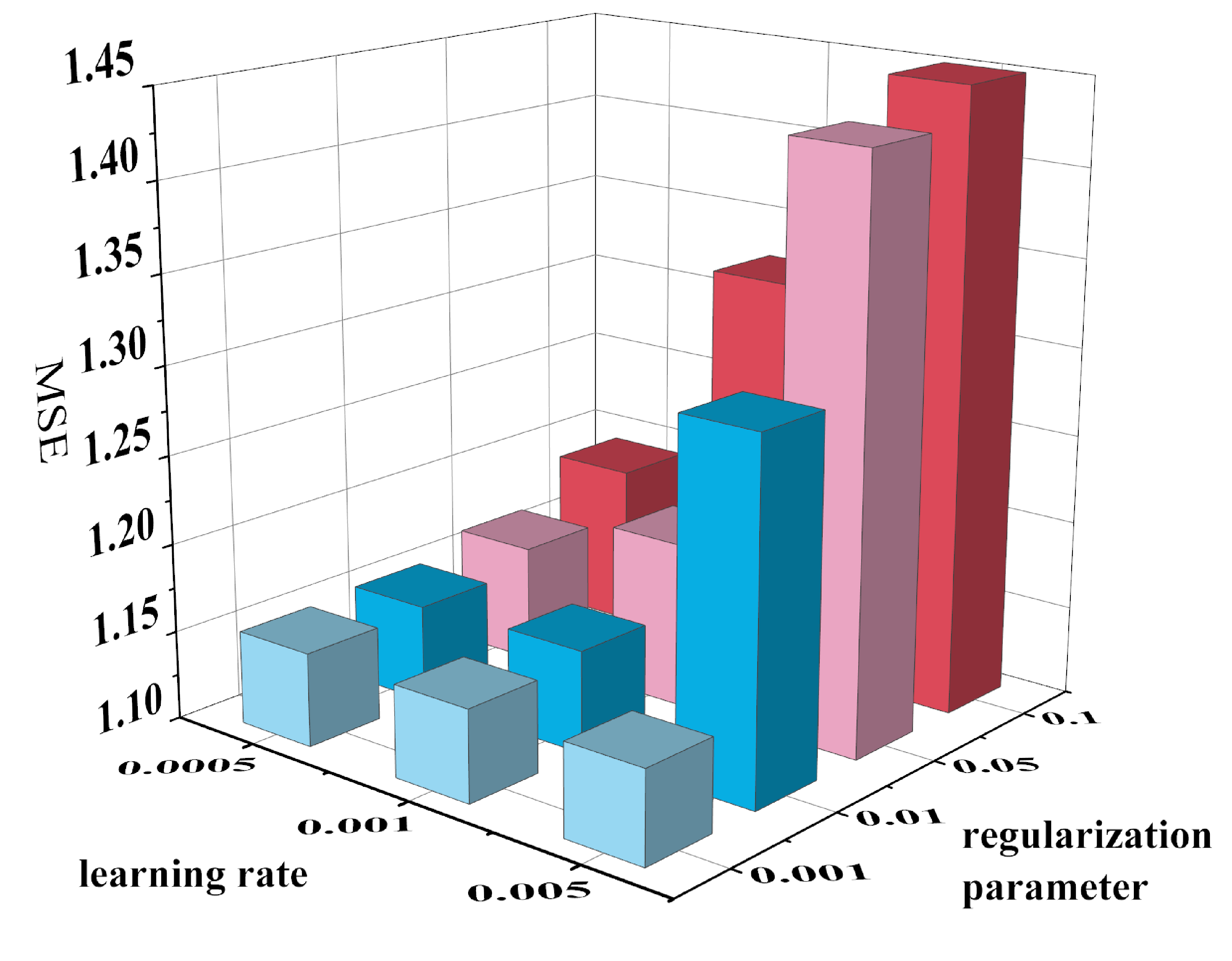}}
    \caption{Sparsity analysis of learning rate $\gamma$ and the regularization parameter $\lambda$ on six datasets.}
    \label{fig:lr_rg_3D}
\end{figure}

We perform experiments to evaluate the sensitivity of APH to its hyperparameters.
Following RGNN, the learning rate $\gamma$ is varied in $[0.0005, 0.001, 0.005]$, and the regularization parameter $\lambda$ is varied in $[0.001, 0.01, 0.05, 0.1]$.
The experiment results are demonstrated in Figure~\ref{fig:lr_rg_3D}, which shows the impact of hyperparameters $\gamma$ and $\lambda$ on six datasets.
We can see that APH likes a small regularization parameter and a big learning rate.
It achieves the best value on most datasets while $\gamma=0.005$ and $\lambda=0.001$.

\begin{figure} 
    \centering
    \includegraphics[width=6cm]{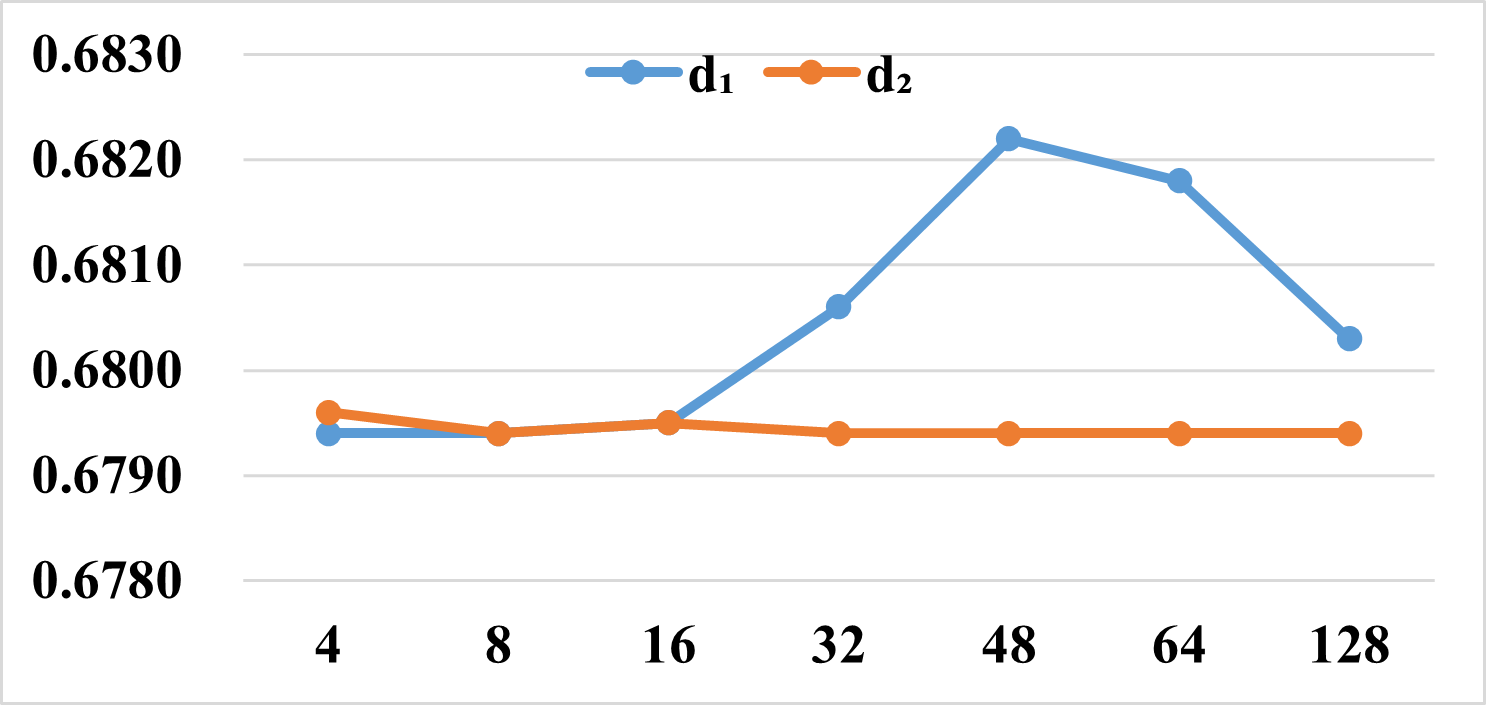}
    \caption{MSE of APH with various dimensions on Music dataset.}
    \vspace{-0.5cm}
    \label{fig:exp_dim}
\end{figure}

\subsubsection{The effect of embedding dimension}
We perform experiments to evaluate the sensitivity of APH to its hyperparameters.
The dimension of the semantic space $d_1$ and the hidden space of MLP $d_2$, are varied in $\{4, 8, 16, 32, 48, 64, 128. \}$
To reduce the computing cost, when we verify the impact of $d_1$, we set $d_2=8$, and that has the same setting for the verification of $d_2$.
Figure~\ref{fig:exp_dim} shows the MSE results on the Music dataset. 
We can see that APH achieves the best performance when $d_1=8$ and $d_2=8$.
On the other five datasets, the MSE results show no significant difference in various dimensions settings.
In APH, users and items are represented not only by their IDs' embedding but also by aggregating their aspect-sentiment graphs.
Thus, APH has great power to represent users and items by little dimension.
For more experiments, please refer to the supplementary materials.

\subsection{Results of extracted aspect}

\begin{figure}
    \centering
    \subfigure[Music]{\includegraphics[width=4cm]{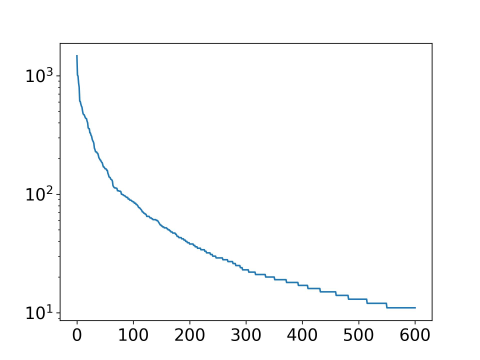}}
    \subfigure[Office]{\includegraphics[width=4cm]{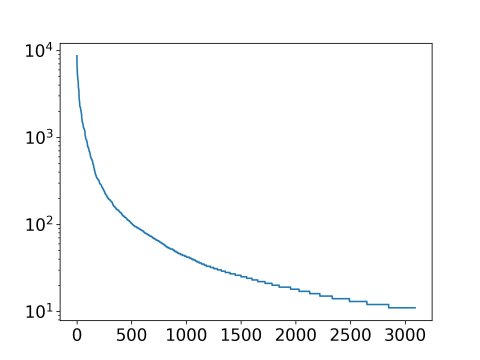}}
    \subfigure[Toys]{\includegraphics[width=4cm]{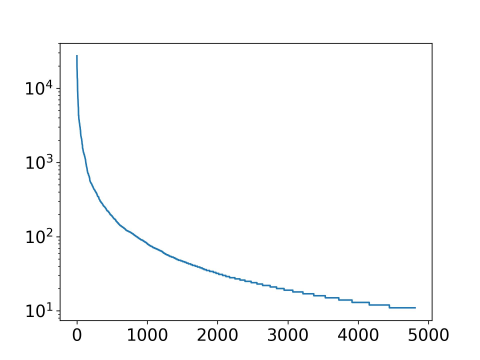}}
    \subfigure[Games]{\includegraphics[width=4cm]{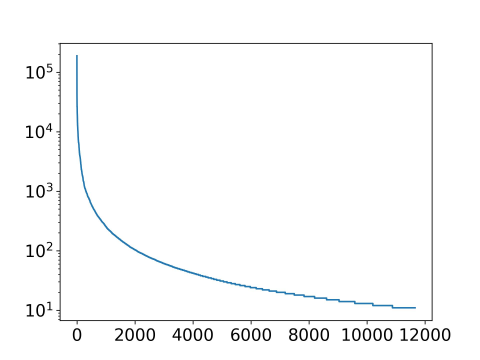}}
    \subfigure[Beauty]{\includegraphics[width=4cm]{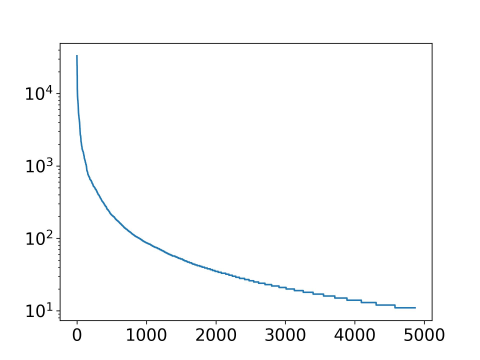}}
    \subfigure[Yelp]{\includegraphics[width=4cm]{log_yelp2.png}}
    \caption{Aspect distribution in six datasets.}
    \label{fig:Aspect_distribution_full}
\end{figure}

Our method aggregates explicit aspects to represent users and items and fuses them to make predictions.
The aspects are extracted by an unsupervised method, discussed in section~\ref{sec:Aspect_graph_constractor_app}.
The number of aspects is smaller than that of items, and the number of quadruples is bigger than that of ratings.
We also give the distribution of aspects in Figure~\ref{fig:Aspect_distribution_full}.
The aspect distributions of all datasets are long-tail distributions.

\end{document}